\documentclass[journal]{IEEEtran}

\ifCLASSINFOpdf

\else

\fi

\usepackage{amssymb}
\usepackage{amsmath, amsthm, amsfonts}
\usepackage{amsfonts} 
\usepackage{graphicx} 
\usepackage{algorithm}
\usepackage{algorithmic}
\usepackage[update,prepend]{epstopdf}
\usepackage{cite}
\usepackage{color}
\usepackage{soul}
\usepackage[caption=false]{subfig}
\usepackage{etoolbox}
\usepackage{caption}
\usepackage{comment}
\usepackage{balance}
\makeatletter
\patchcmd{\@makecaption}
  {\scshape}
  {}
  {}
  {}
\makeatletter
\patchcmd{\@makecaption}
  {\\}
  {.\ }
  {}
  {}
\makeatother

\begin{document}
%

\title{ViLDAR -- Visible Light Sensing Based \\ Speed Estimation using Vehicle's Headlamps}

\markboth{Submitted to IEEE Transactions on Vehicular Technology,~Vol.~X, No.~X, XX~2018}{Shell \MakeLowercase{\textit{et al.}}: Bare Demo of IEEEtran.cls for Journals}

\author{Hisham~Abuella,~\IEEEmembership{Student~Member,~IEEE,} Farshad~Miramirkhani,~\IEEEmembership{Student~Member,~IEEE,}
Sabit~Ekin,~\IEEEmembership{Member,~IEEE,}~Murat~Uysal,~\IEEEmembership{Senior~Member,~IEEE}, and~Samir~Ahmed

\thanks{This paper was presented in part at the IEEE 38th Sarnoff Symposium in Newark, NJ, Sept. 2017.}

\thanks{H.~Abuella and S.~Ekin are with the School of Electrical and Computer Engineering, Oklahoma State University, Oklahoma, USA (e-mail:~hisham.abuella@okstate.edu,~sabit.ekin@okstate.edu).}
\thanks{F.~Miramirkhani and M.~Uysal are with the Department of Electrical and Electronics Engineering, Ozyegin University, Istanbul, Turkey, (e-mail:~farshad.miramirkhani@ozu.edu.tr, murat.uysal@ozyegin.edu.tr).}
\thanks{S.~Ahmed is with the School of Civil and Environmental Engineering, Oklahoma State University, Oklahoma, USA (e-mail:~sahmed@okstate.edu).}}

\maketitle

\begin{abstract}

The introduction of light emitting diodes (LED) in automotive exterior lighting systems provides opportunities to develop viable alternatives to conventional communication and sensing technologies. Most of the advanced driver-assist and autonomous vehicle technologies are based on Radio Detection and Ranging (RADAR) or Light Detection and Ranging (LiDAR) systems that use radio frequency or laser signals, respectively. While reliable and real-time information on vehicle speeds is critical for traffic operations management and autonomous vehicles safety, RADAR or LiDAR systems have some deficiencies especially in curved road scenarios where the incidence angle is rapidly varying. In this paper, we propose a novel speed estimation system so-called the Visible Light Detection and Ranging (ViLDAR) that builds upon sensing visible light variation of the vehicle's headlamp.
We determine the accuracy of the proposed speed estimator in straight and curved road scenarios. We further present how the algorithm design parameters and the channel noise level affect the speed estimation accuracy. For wide incidence angles, the simulation results show that the ViLDAR outperforms RADAR/LiDAR systems in both straight and curved road scenarios.

A provisional patent (US\#62/541,913) has been obtained for this work.

\end{abstract}


\IEEEpeerreviewmaketitle

\section{Introduction}


\IEEEPARstart{I}n recent years, most vehicle manufacturers have started equipping their vehicles with daytime running lights (DRLs) by using light emitting diode (LED) headlamps because of their advantages such as long life time, energy efficiency, and short rise time. 
Research results show that DRLs can decrease the crash rate by up to 28\% for multi-vehicle and pedestrian crashes \cite{koornstra1997safety, DRL, paine2003review}. LEDs can be modulated at very high frequencies without any adverse effects on illumination level. Therefore, the idea of using visible light communication (VLC) in the vehicle-to-X (V2X) communication to replace radio frequency (RF) based standards (i.e., 802.11p) has been already proposed in \cite{VLC_ieee802_11_p,Study1_USING_VLC_V2V_Comm, Study3_USING_VLC_V2V_Comm, Study4_USING_VLC_V2V_Comm,Study5_USING_VLC_V2V_Comm, Farshad_ref14, Study7_USING_VLC_V2V_Comm}. 



Another potential application of LED headlamps is visible light sensing (VLS) for speed estimation which will be the focus of this paper.
Transportation agencies around the world measure vehicle speed for a number of reasons. Average speed is used to measure the quality of traffic operations and the reliability of travel time. The distribution of vehicle speeds is used to determine the traffic performance, examination of highway consistency and safety. Speed enforcement programs aim to lower the number and severity of traffic crashes. Therefore, transportation traffic professionals are always looking for technological advancements to enhance speed measurement systems. 

\textbf{RA}dio \textbf{D}etection \textbf{A}nd \textbf{R}anging (RADAR) system is a popular method for vehicle speed estimation. Depending on the application and scenario, a RADAR can be used for either detecting a moving object or estimating its speed \cite{RADAR_principles_Book}. A similar system that uses the same principle of the RADAR but works on a different part of the electromagnetic radio spectrum is the \textbf{Li}ght \textbf{D}etection \textbf{A}nd \textbf{R}anging (LiDAR)). RADAR and LiDAR systems estimate the speed by detecting the change in frequency and travel time of the reflected signal. Among various challenges, the beam-width and angle-of-arrival are the most critical ones and have high impact on the estimation accuracy of these systems \cite{Doppler_RADAR_Gun_Performance2}. Indeed, narrow beam-width is required for accurate speed estimations. Furthermore, RADAR or LiDAR warning devices/systems are commercially available and can be easily acquired by drivers \cite{Avoiding_RADAR}. These devices detect the presence of RADAR/LiDAR (also known as speed gun or RADAR gun) and warn the driver to avoid traffic fines for speeding. More information about state-of-the-art RADARs and LiDARs' disadvantages and limitations is provided in details in\cite{Improving_police_radar}.


\begin{figure*}[h!]
\centering
     \subfloat[]{%
       \includegraphics[width=0.52\textwidth]{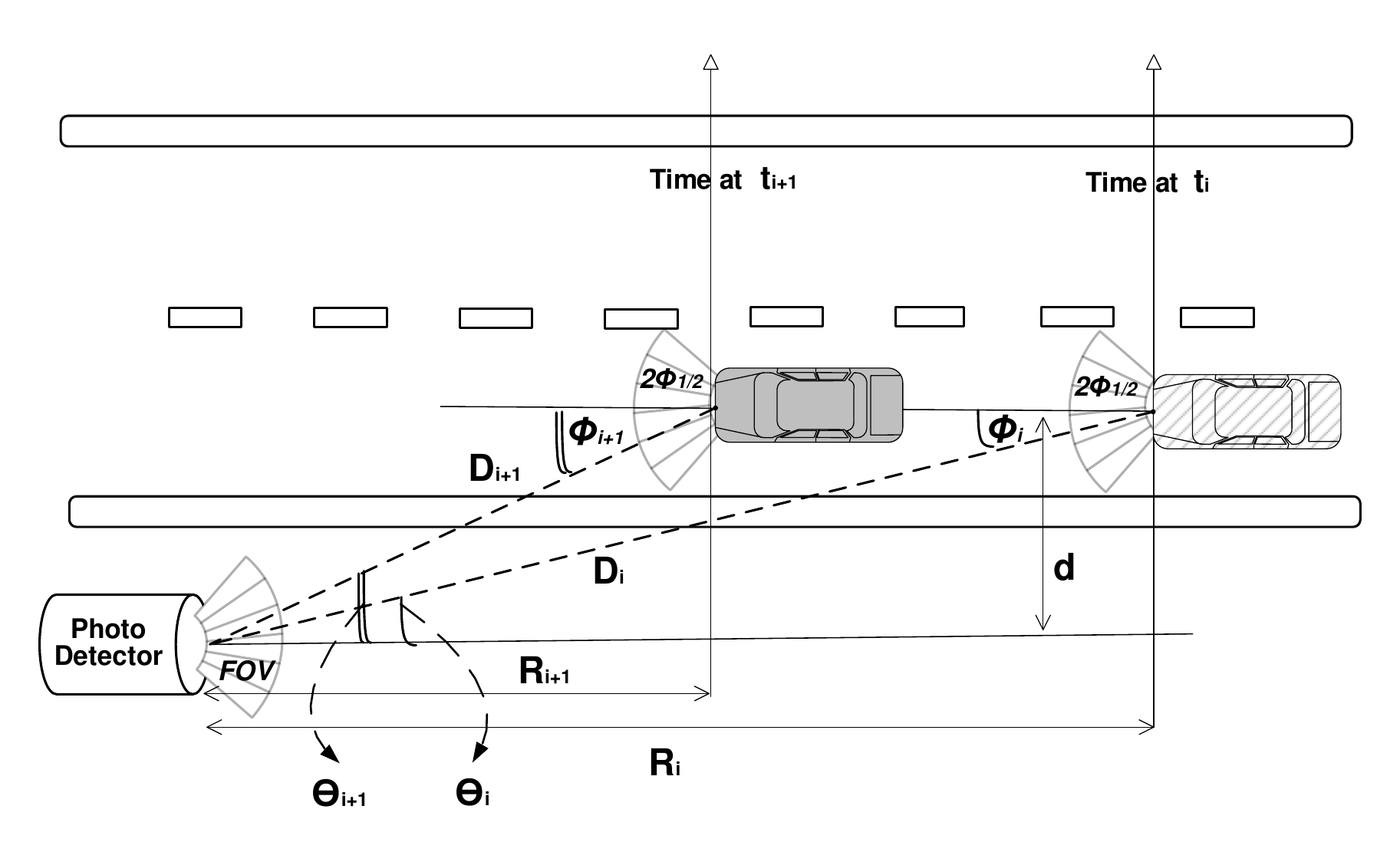}
     }
     \subfloat[]{%
       \includegraphics[width=0.4\textwidth]{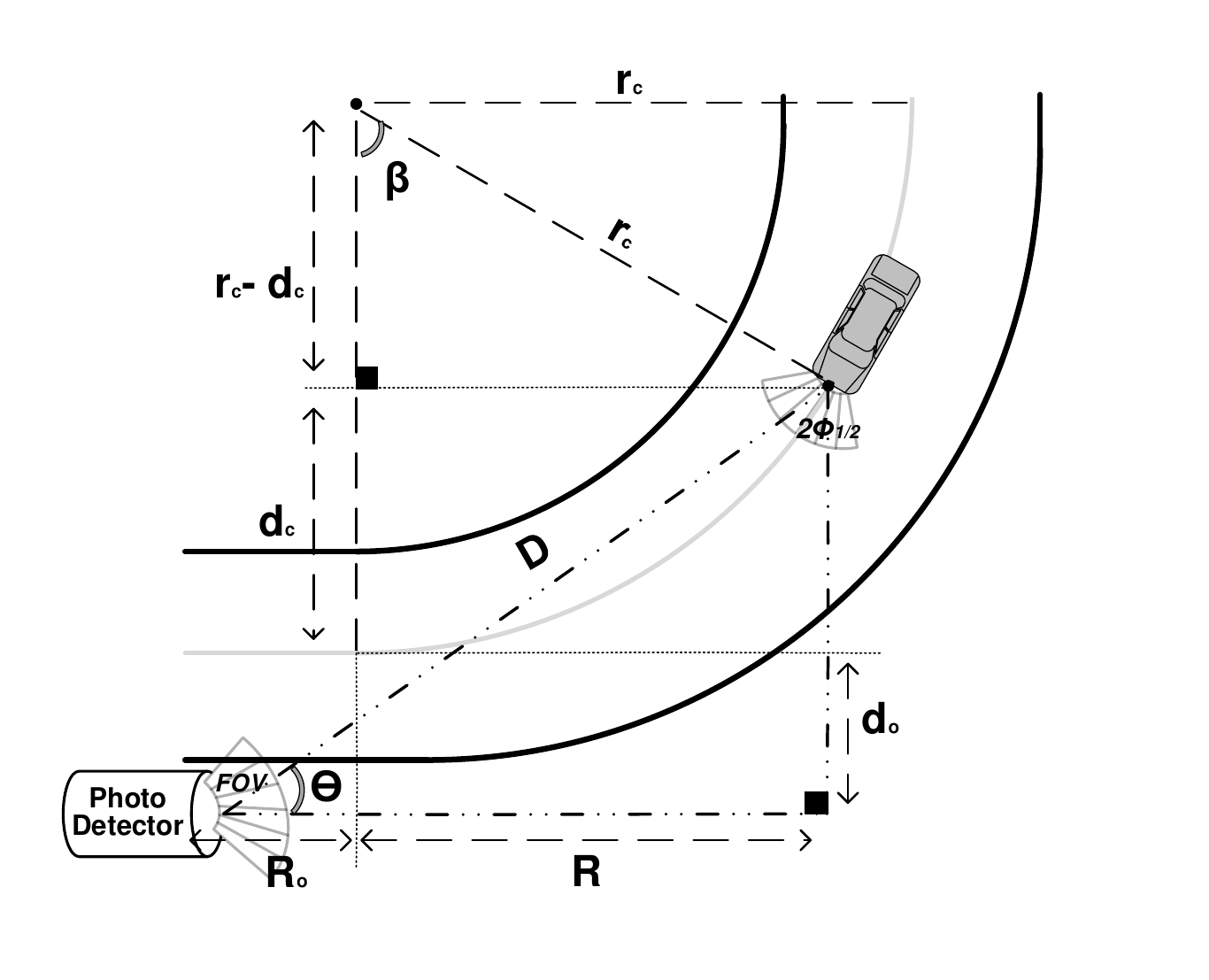}
     }
     \caption{ViLDAR system model (a) straight road and (b) curved road.}
     \label{fig:System_Model}
   \end{figure*}

In this paper, we introduce a VLS based speed estimation system that uses received light power (intensity) variations of an approaching vehicle's headlamps. We termed the system as \textbf{Vi}sible \textbf{L}ight \textbf{D}etection \textbf{A}nd \textbf{R}anging (ViLDAR)\footnote{The patent is pending: H.  Abuella,  S.  Ekin,  and  M.  Uysal,  “System and method for speed estimation, in vehicles,”  US Patent App. No. 62/541,913.}, where the vehicle's headlamp acts as a transmitter. As long as the vehicle light is in ViLDAR’s field of view (FOV) (similar to beam-width in RADAR), the irradiance angle (angle of arrival) as low impact on estimation accuracy. To the extent of our knowledge, the concept of utilizing headlamps' received power level for vehicle speed estimation has not been discussed in the literature previously.


The main contributions of this study are given as follows:
\begin{itemize}
\item  A VLS-based detection and ranging system using vehicle's headlamp is presented. 

\item  The ViLDAR system performance is simulated based on channel coefficients generated from an advanced ray tracing tool to mimic the realistic and physical-based visible light channel models.

\item  The performance of ViLDAR system in both curved and straight road scenarios is investigated and compared with theoretical  performance of RADAR system in ideal environment conditions.

\end{itemize}

The advantages of the ViLDAR system are summarized as follows:
\begin{itemize}
\item  It is a low-power ranging and detection system that reflects on good battery efficiency.
 
\item  It has less noise and pathloss compared to two-way model.

\item  It performs better in a large incident angle and in scenarios where the incident angle is varying fast, i.e., curved road scenarios.

\item  Size and weight of ViLDAR system will be smaller and lighter than current handheld law enforcement RADAR guns.

\item Unlike RADAR guns, the ViLDAR system cannot be detected by drivers when used by law enforcement officers.

\item Since ViLDAR uses the visible light, it is not affected by RF interference and does not cause any interference to other RF systems.
\end{itemize}

The rest of the paper is organized as follows. First, we present the system model in Section~\ref{sec:sys_model}. Channel models under considerations are discussed in Section~\ref{sec:Channel_model}. The speed estimation methods are presented in Section~\ref{sec:Speed_Estimation_Algorithms}.
 The numerical results are presented in Section~\ref{sec:Sim_Results}, while conclusions are drawn in Section VI.

\section{System Model}\label{sec:sys_model}


The system model is depicted in Fig.~\ref{fig:System_Model} where both straight and curved road deployments are illustrated. In Fig.~\ref{fig:System_Model}.a, $\theta$ and $d$ are the incidence angle and the vertically projected distance between the vehicle and the photodetector (PD), respectively. Likewise, $R_i$ and $D_i$ are the horizontally projected distance and the actual distance between the vehicle and the PD at time $t_i$, respectively. To avoid confusion, $D$ and $R$ are called as distance and range for the rest of the paper. Moreover, the subscript $i$ and $i+1$ stand for the values of the corresponding parameter at time instances $t_i$ and $t_{i+1}$, respectively.

In Fig.~\ref{fig:System_Model}..b, $r_c$ is the curvature radius of the road, $\beta$ is the angle that is changing with respect to the angular velocity of the car $w = \frac{V}{r_c} $ and $V$ is the vehicle speed. In this scenario, both the horizontal distance from ViLDAR to vehicle $d_c$ and vertical distance $R$ change with respect to $\beta$. In addition, $d_o$ and $R_o$ are the horizontal and vertical distances between ViLDAR and the end of the curved road, respectively.

The PD in ViLDAR system takes new measurements with a certain sampling time (i.e., different $D$ and $R$ from the PD). The received power increases as the vehicles approach the PD. Given the channel model, the speed estimation problem can be interpreted as designing an estimator to obtain the slope of the received power (see Fig. 5); hence, estimate the speed of the vehicle. Lastly, We assume that the vehicle moves with constant velocity. It is further assumed that 1) vehicle's LED transmits a constant power (luminance), 2) only a single\footnote{Depending on application, if needed, the position of the PD can be changed to focus on a single lane to perform speed estimation of a single vehicle.} vehicle approaches in the duration of measurements, i.e., certain distance between vehicles is assumed, 3) the ViLDAR's FOV\footnote{This assumption can be realized by using some optical lens with the PD.} is assumed to be 70$^{\circ}$, and 4) difference between the two headlamps in the vehicle is small, hence it is assumed that vehicle has one transmitter.

\section{Channel Modeling}
\label{sec:Channel_model}
While VLC has been studied intensively in the context of indoor communications \cite{VLC_State_of_Art,VLS_REF}, its application to vehicular networking is relatively new \cite{Farshad_ref9,Farshad_ref10,Farshad_ref11}. Earlier works on infrastructure-to-vehicle (I2V) links (i.e., from traffic light to vehicle) \cite{Farshad_ref12,Farshad_ref13} build upon the line-of-sight (LOS) channel model originally proposed for the indoor LED light sources with Lambertian pattern. However, such a model is not applicable to automotive low-beam and high-beam headlamps with asymmetrical intensity distributions. To address this, a piecewise Lambertian channel model was proposed in \cite{Farshad_ref14} to reflect the asymmetrical intensity distribution of scooter taillight. Measured intensity distribution patterns were further used in \cite{Farshad_ref15,Farshad_ref16,Farshad_ref17} to accurately reflect the asymmetrical structure of automotive lighting in vehicle-to-vehicle (V2V) channel modeling.

The reflections from road surface might impact the vehicular VLC system performance. The reflectance of road surface depends on its nature and physical state. In \cite{Farshad_ref15}, Lee \textit{et al.} utilized Monte Carlo ray tracing to obtain channel delay profiles for V2V, I2V and vehicle-to-infrastructure (V2I) links for a road surface with fixed reflectance value. In \cite{Farshad_ref16,Farshad_ref17}, Luo \textit{et al.} proposed a geometric V2V VLC channel model based on the measured headlamp beam patterns and a road surface reflection model. The link BER performance was investigated for the clean and dirty headlamps in a wet and dry road surface. In \cite{Farshad_ref20}, Elamassie \textit{et al.} carried out a comprehensive channel modeling study to quantify the effect of rain and fog on a V2V link with a high-beam headlamp acting as the transmitter. Taking advantage of advanced ray tracing features, they developed a path loss model for V2V link as a function of distance under different weather conditions.

In this section, we explore two path loss models for ViLDAR system under consideration. As hypothetical case, we first consider Lambertian channel model. Then, we adopt ray-tracing approach \cite{Farshad_ref20} to propose more realistic channel model taking into account several practical constrains such as the asymmetrical pattern of headlamp and street lights, reflections from road surfaces and weather conditions.

 \subsection{Lambertian Channel Model}
 In this model, the power-distance relation is given by\cite{VLC_First_channel_Model}
   \begin{equation}
   \label{eq:Lambertian_Channel1}
                  P_r(t)=\frac{(n+1)A_R P_t}{2\pi [D(t)]^{\gamma}}\cos^n(\phi(t))\cos(\theta(t)) , \forall\theta(t)< \phi_{1/2},
  \end{equation}
where $P_t$ is the transmitter power and $A_R$ is the optical PD size.  $ \phi(t) $ and $ \theta(t) $ are irradiance and incidence angles at time $t $, respectively.
In addition, $\phi_{1/2}$ is the semi-angle at half-power of the LED, and $n$ is the order of the Lambertian model and is given by
\begin{equation}
n=-\frac{\ln(2)}{\ln(\cos\phi_{1/2})}.
\end{equation}

Assuming same heights for both transmitter and receiver, we have
     \begin{equation}
     \label{eq:Theta_PHI}
                   \theta(t) = \phi(t) ,\forall t,
  \end{equation}
 where  $ 0 <\theta(t) < \phi_{1/2} $. Using~\eqref{eq:Theta_PHI},~\eqref{eq:Lambertian_Channel1} can be further simplified as
\begin{equation}
 \label{eq:Lambertian_Channel2}
 P_r(t)=\frac{(n+1)A_R}{2\pi [D(t)]^{\gamma}}\cos^{n+1}(\theta(t)).
 \end{equation}

Finally, in order to derive $P_r(t)$ in terms of $D(t)$, we further simplify \eqref{eq:Lambertian_Channel2} by defining a constant as  

\begin{equation}
 \label{eq:C_Lambertian_Channel}
C = \frac{(n+1)A_R}{2\pi}.
\end{equation}

Using the fact that $\cos(\theta)=\frac{\sqrt{[D(t)]^2 -d^2}}{D(t)}$, the resulting expression yields
     \begin{equation}
   \label{eq:Lambertian_Channel3}
                  P_r(t)=\frac{C}{[D(t)]^{\gamma}}\left (\frac{\sqrt{[D(t)]^2 -d^2}}{D(t)}\right )^{n+1}.
  \end{equation}

\subsection{Simulated Channel Model}
\label{sec:Realistic_Channel_model}

\begin{figure*}[h!]
\centering
     \subfloat[]{%
       \includegraphics[clip,trim=5 0 5 20, width=0.522\textwidth]{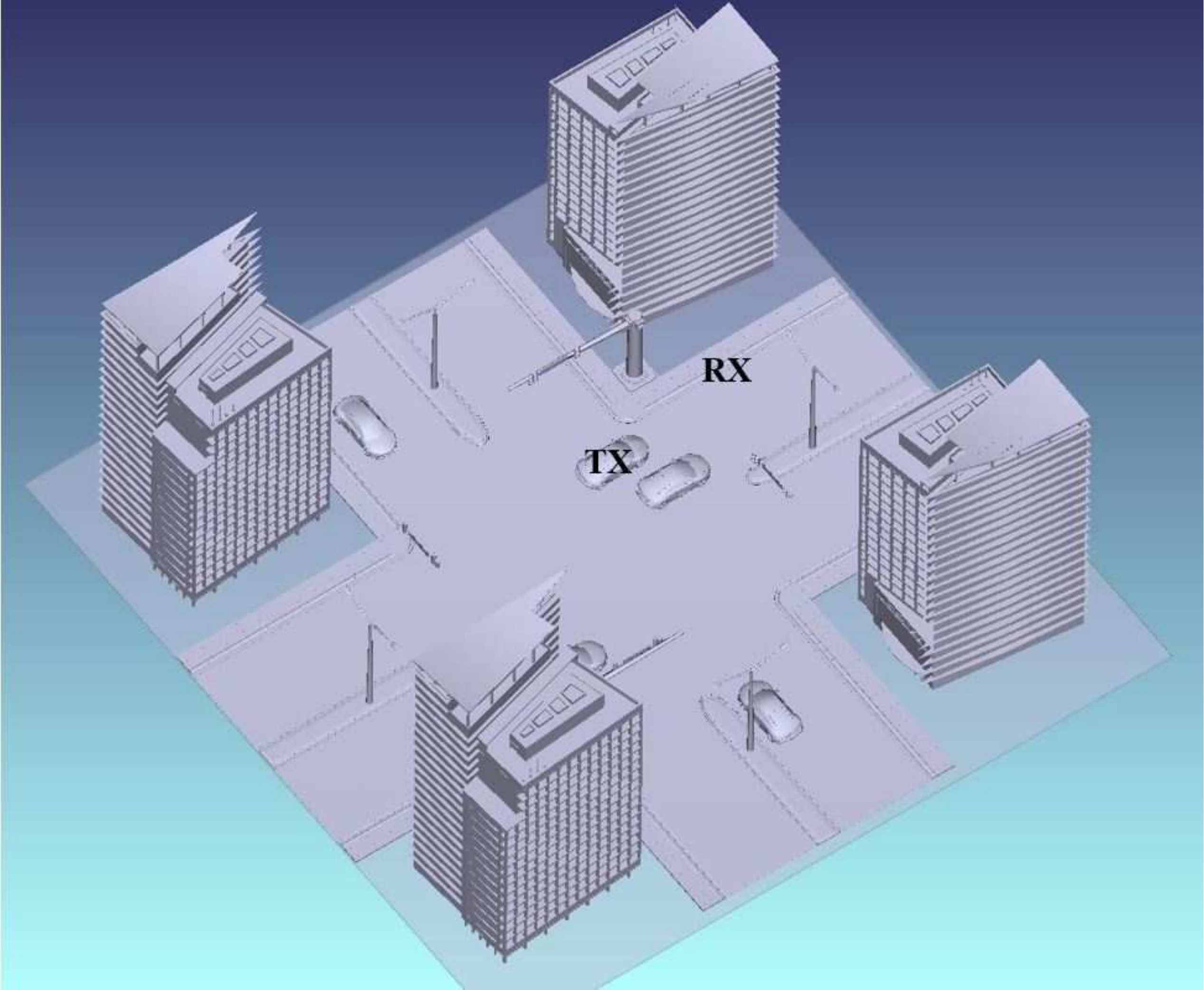}
     }
     \subfloat[]{%
       \includegraphics[clip,trim=5 0 5 11, width=0.282\textwidth]{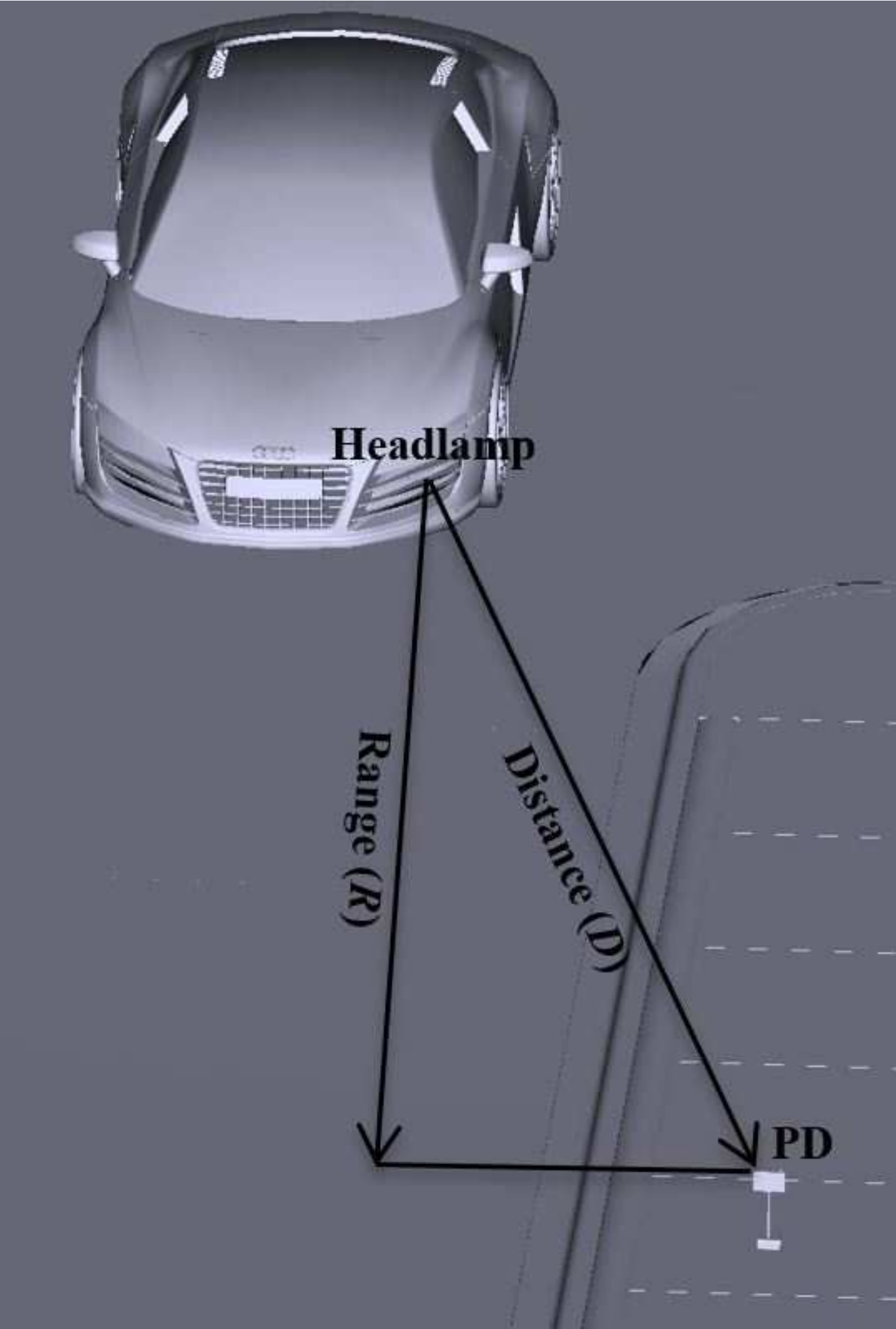}
     }
     \caption{(a) Scenario under consideration and (b) transmitter and receiver close-up in the 3-D ray tracer.}
     \label{fig:Zemax_setup}
   \end{figure*}

We use a similar channel modeling methodology based on Zemax\textsuperscript{\textregistered} as in \cite{Farshad_ref20}. We first construct the simulation platform of the outdoor environment integrating the CAD models of building, vehicles and any other objects within. We further specify the type of object surface materials (coating) and the types of reflections, i.e., purely diffuse, specular and mixed reflections. The specific type of reflection is defined by “scatter fraction” parameter. Mie scattering is further used to model clear weather conditions \cite{Farshad_ref23}. ``Bulk scatter'' method in the software allows providing the input parameters ``particle index'' (the refractive index of particles), ``size'' (the radius of the spherical particles) and ``density'' (the density of particles).

After we create the simulation environment, we use the built-in ray tracing function to determine the CIR. The non-sequential ray tracing tool generates an output file, which includes the detected power and path lengths from source to PD for each ray. We import this file to Matlab\textsuperscript{\textregistered} and, using this information, we can express the CIR as
     \begin{equation}
   \label{eq:Channel_responce}
                  h(t)=\sum^{N_r}_{i=1}P_i\delta(t-\tau_i),
  \end{equation}
where  $P_i$ is the power of the $i$th ray, $\tau_i$ is the propagation time of the $i$th ray, $\delta(t)$ is the Dirac delta function and $N_r$ is the number of rays received at the PD. 

The received optical power is given as $P_r(t)=P_t-PL$ where $P_t$ is the transmit optical power and $PL$ is expressed as \cite{Farshad_Ref3}
    \begin{equation}
   \label{eq:Path_Loss}
                 PL= 10 \log_{10}\left ( \int_0^\infty  h(t)dt   \right ).
  \end{equation}
In the simulated channel model (similar to the RF path loss model \cite{mmWave_Rappaort_Book}), the power-distance relation can be given by
  \begin{equation}
  \label{eq:Realistic_Channel}
                  P_r(t)=K[D(t)]^{-\gamma},    \forall~D(t) > 1,
  \end{equation}
where $P_r(t)$ is the power level received from the vehicle at time $t$, and $K$ is a constant that represents all the gains and the transmitted power. $\gamma$ is the channel path-loss exponent, which usually depends on the channel environment, and $D(t)$ is the distance between the vehicle and the ViLDAR at time $t$.

\begin{figure}[t]
\centering
     \subfloat[]{%
       \includegraphics[width=0.47\textwidth]{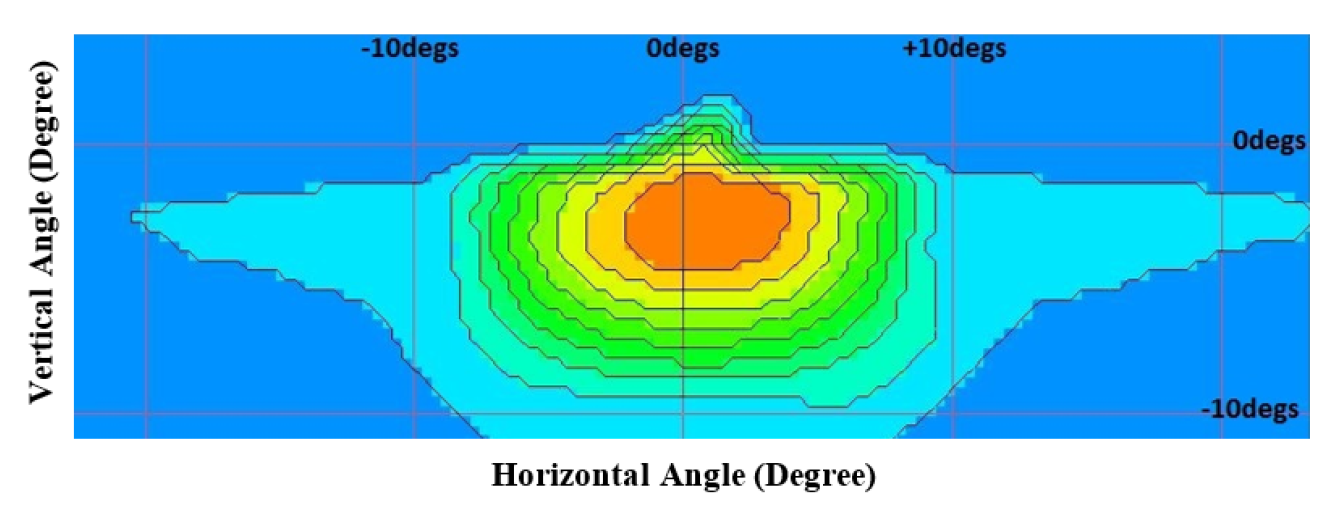}
     }
     \hfill
     \subfloat[]{%
       \includegraphics[width=0.38\textwidth]{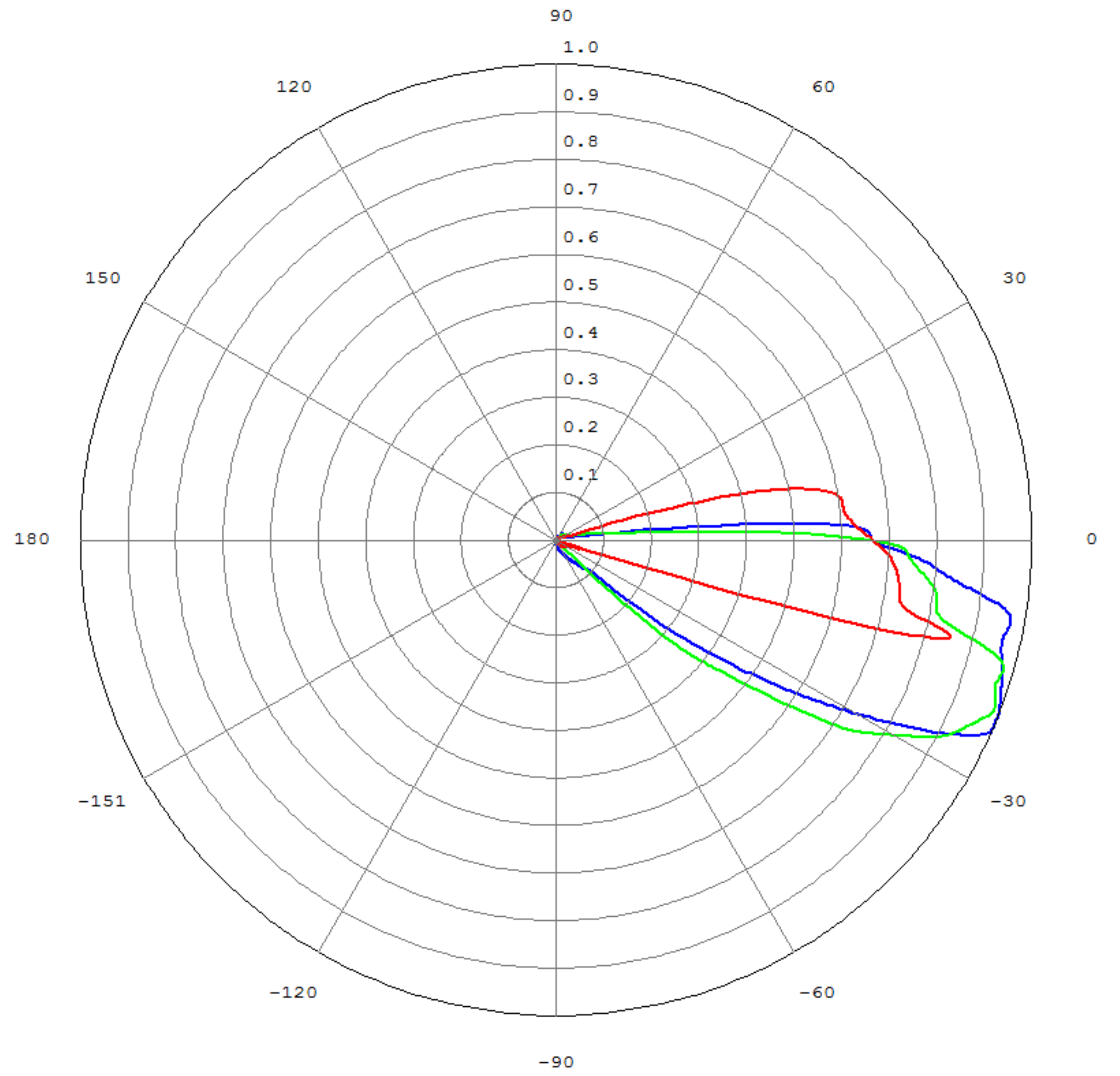}
     }
     \caption{(a) Spatial distribution and (b) relative intensity distribution (C0$^{\circ}$-C180$^{\circ}$, C90$^{\circ}$-C270$^{\circ}$ and C135$^{\circ}$-C315$^{\circ}$ planes are respectively indicated by red, blue and green color) of low-beam headlamp under consideration.}
     \label{fig:Zemax_setup1}
     \vspace{-0.6cm}
   \end{figure}

We consider a scenario shown in Fig.~\ref{fig:Zemax_setup1}. We assume that coating materials of buildings, traffic light poles and street lamp poles are respectively concrete, aluminum metal and galvanized steel metal. The coating material of cars is considered as black and olive green gloss paint. The road type is assumed as R3 with the coating material of asphalt\cite{Farshad_Ref22}. 

We use Philips Luxeon Rebel automotive white LED as the low-beam headlamp with the spatial distribution shown in Fig.~\ref{fig:Zemax_setup1}.a. Due to asymmetrical intensity distribution of luminaire, different cross sections indicated by C0$^{\circ}$-C180$^{\circ}$, C90$^{\circ}$-C270$^{\circ}$ and C135$^{\circ}$-C315$^{\circ}$ planes are shown in Fig.~\ref{fig:Zemax_setup1}.b. The headlamp with total power normalized to unity is placed in the front side of the vehicle as the transmitter and the PD with 1 $\text{cm}^2$ area and FOV of 70$^{\circ}$ is placed on the sidewalk (see Fig.~\ref{fig:Zemax_setup}.b).

We assume that the vehicle moves toward the PD. We obtain the CIRs through all points with 1 meter interdistance over the driving direction of the car for a range of 10 meter. In an effort to obtain $K$ and $\gamma$ values in (\ref{eq:Realistic_Channel}), we apply curve fitting techniques on our calculated path loss in (\ref{eq:Path_Loss}) based on the minimization of root mean square error. The related coefficients $K_{dB}$ and $\gamma$ are presented in Table~\ref{table:Channel_Models_Parameters}. The Lambertian channel model is also included as a benchmark (i.e., hypothetical case).

\begin{figure}[t!]
\includegraphics[width=0.48\textwidth]{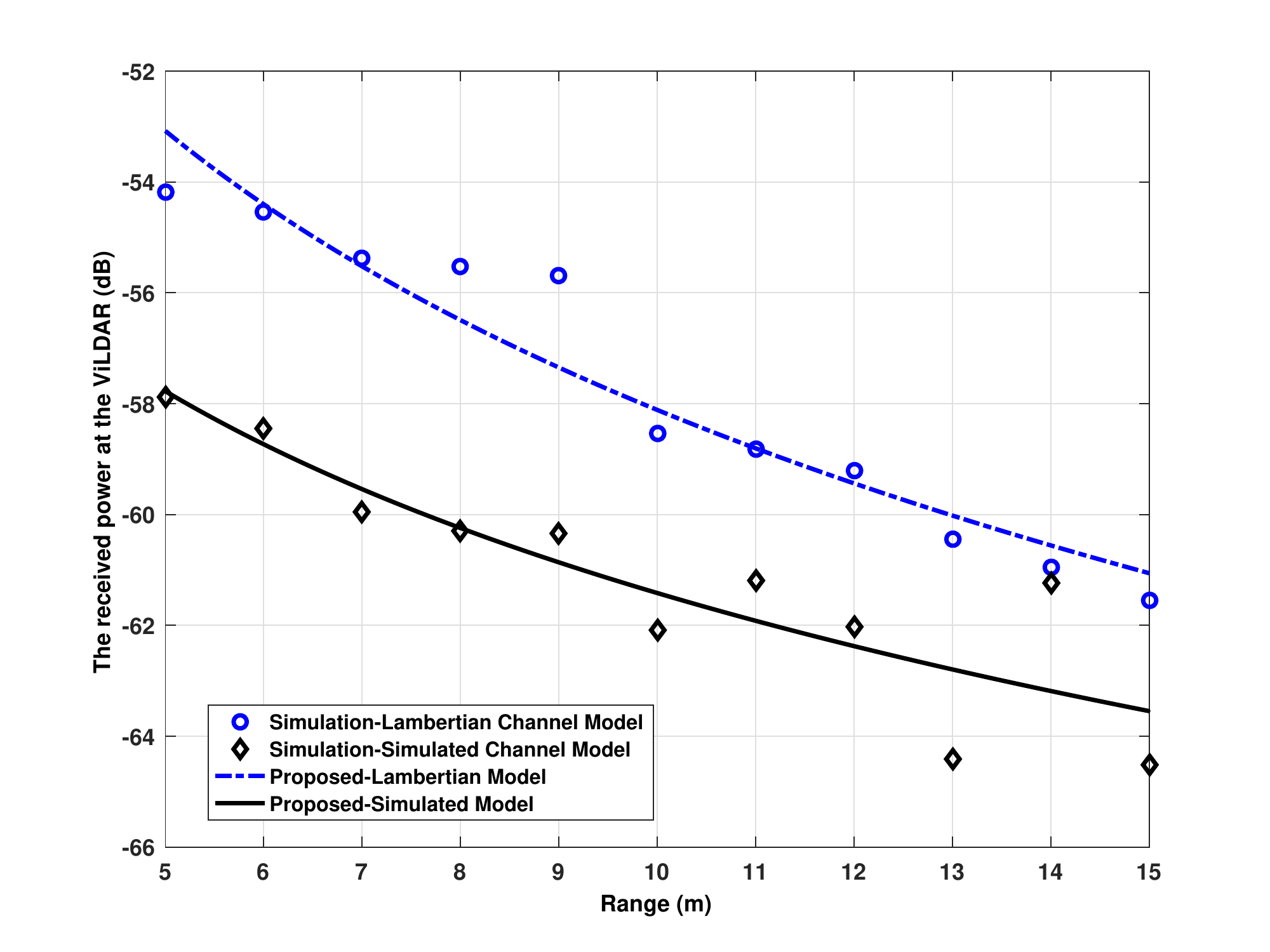}
\centering
\caption{Path Loss versus distance for Lambertian and simulated channel models.}
\label{fig:Channel_Models_Fitting}
\vspace{-0.5cm}
\end{figure}

In Fig.~\ref{fig:Channel_Models_Fitting}, we present the path loss versus distance for the channel models under consideration. It is observed from Fig.~\ref{fig:Channel_Models_Fitting} that the path loss obtained with Lambertian channel model is underestimated with respect to the simulated channel model. This is a result of the fact that in the simulated channel model more reflected rays from the road surface are received.

\begin{figure}[t!]
\includegraphics[width=0.485\textwidth]{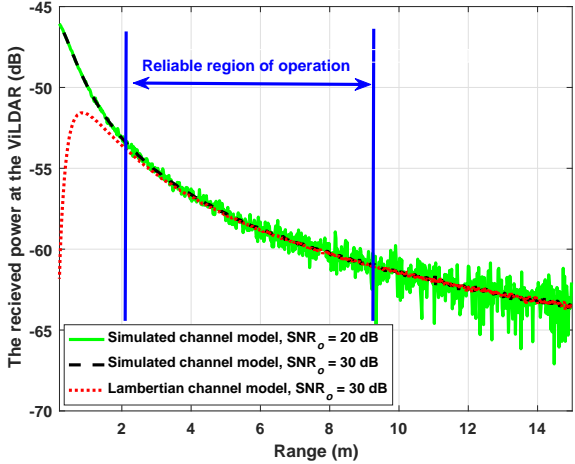}
\centering
\caption{Reliable region of operation in ViLDAR for Lambertian and simulated channel models.}
\vspace{-0.5cm}
\label{fig:Realistic_Lambertian_ChannelModel}
\end{figure}

In Fig.~\ref{fig:Realistic_Lambertian_ChannelModel}, Lambertian and simulated channel models are compared. To have a fair comparison, we use the same parameter values in both channel models. For instance, the constant $C$ in~\eqref{eq:C_Lambertian_Channel} equals to constant $K$ in~\eqref{eq:Realistic_Channel} and with the same $\gamma$ for path-loss exponent\footnote{Both $K_{dB}$ and $\gamma$ values are estimated using a ray tracing simulation explained in Section~\ref{sec:Channel_model}. They are environment dependent.}. In order to show the impact of noise level, different \textit{initial} signal-to-noise-ratios (SNR$_o$) are used in the simulations (20 dB and 30 dB ).

\begin{table}[t!]
\fontsize{11}{9}
\centering
\caption{Channel parameters for Lambertian and simulated channel models.}
\label{table:Channel_Models_Parameters}
 \begin{tabular}{||c | c c||} 
 
 \hline
 &\footnotesize $K_{dB}$ & $\gamma $  \\ 
 \hline\hline
 Lambertian Channel Model & -41.39 & 1.673 \\ 
 \hline
 Simulated Channel Model & -49.32 & 1.210 \\
 \hline
\end{tabular}
\vspace{-0.4cm}
\end{table}

Initially at time 0, and as the vehicle advances to the PD, distance decreases, ViLDAR takes new measurements. As predicted, the received power increases as the vehicle approach the PD in both models. Furthermore, although the estimation can be performed from all the received power levels, high accuracy of speed estimation can be obtained in certain region, which is shown as the reliable region of operation in Fig.~\ref{fig:Realistic_Lambertian_ChannelModel}. 

\section{Speed Estimation}\label{sec:Speed_Estimation_Algorithms}

In this section, we present the speed estimation algorithms by using the channel models presented in preceding section for two different road scenarios. 

\subsection{Estimation in Straight Road Scenario}
First, we consider the simulated channel model presented in (\ref{eq:Realistic_Channel}).
Under the assumption of constant speed during the estimation period, $D(t)$ can be expressed in terms of speed and time as
   \begin{equation}
    \label{eq:Distance_Time_V_Ro}
                  D(t) =  \sqrt{d^2 + \left(R_o - Vt \right)^2} ,
  \end{equation}
where $R_o$ is the initial range between the vehicle and the ViLDAR. Then, replacing~\eqref{eq:Distance_Time_V_Ro} in~\eqref{eq:Realistic_Channel} yields
   \begin{equation}
   \label{eq:Linear_Model_LS1}
                  P_r(t) = K(d^2 + \left(R_o - Vt \right)^2)^{-\frac{\gamma}{2}}.
  \end{equation}

Rearranging~\eqref{eq:Linear_Model_LS1} results in linear model ($\textbf{\textit{y}} = V\textbf{\textit{x}} + R_o$), we have 
\begin{equation}
   \label{eq:Linear_Model_LS2}
     \sqrt{\left (\frac{P_r(t)}{K}\right )^{-\frac{2}{\gamma}} - d^2} = -V t + R_o,
  \end{equation}
where $\textbf{\textit{y}}$ can be considered as $n$-samples vector of $\sqrt{\left (\frac{P_r(t)}{K}\right )^{-\frac{2}{\gamma}} - d^2}$, and each value in $\textbf{\textit{y}}$ corresponds to a value in vector $\textbf{\textit{x}}$ ($n$-samples vector of negative value of time).
 Then, it can be expressed in vector-domain as
\begin{equation}
\textbf{\textit{y}}= [\textbf{\textit{x}}, \textbf{1}][V , R_o]^T,
\end{equation}
 where $\textbf{1}$ is a vector of $1$'s with size $(n,1)$.
 Then, by letting $\textbf{\textit{b}}=[V , R_o]^T$ with size $(2,1)$
  and $\textbf{A}=[\textbf{\textit{x}} , \textbf{1}]$ with size $(n,2)$, we have
 \begin{equation}
 \textbf{\textit{y}}=\textbf{A}\textbf{\textit{b}}.
 \end{equation}
 
 $V$ and $R_o$ can be readily estimated by using the least square (LS) inverse formula as
  \begin{equation}
   \label{eq:Linear_Model_LS_Final}
     \textbf{\textit{b}}= [\textbf{A}^T \textbf{A}]^{-1} \textbf{A}^T \textbf{\textit{y}}.
  \end{equation}




On the other hand, estimating the speed when using the Lambertian channel model is as follows. First, we update the LS model using the Lambertian channel model as:
  \begin{equation}
  \label{eq:Lambertian_Channel_Proof1}
                  P_r(t)=\frac{(n+1)A_R P_t}{2\pi [D(t)]^{\gamma}}\cos^{(n+1)}(\theta(t)),
  \end{equation}
setting $K = \frac{(n+1)A_R P_t}{2\pi}$, which is a constant value, one can get
  \begin{equation}
  \label{eq:Lambertian_Channel_Proof2}
                  P_r(t)=K [D(t)]^{-\gamma}\cos^{(n+1)}(\theta(t)).
\end{equation}
For $ \cos(\theta(t)) = 1$, the expression in~\eqref{eq:Lambertian_Channel_Proof2} reduces to~\eqref{eq:Realistic_Channel}. To obtain the speed, a similar methodology is followed as in the simulated channel model. To minimize redundancy, we are not going to repeat the same derivation. however, the main difference is the constant parameter $K$. Hence, the same formula in~\eqref{eq:Linear_Model_LS_Final} is applied to estimate the speed in case of using the Lambertian channel model.

\subsection{Estimation in Curved Road Scenario}

As shown in Fig.~\ref{fig:System_Model}.b, the curved road scenario has different setup and parameters than straight road scenario. Hence, we use a different method to estimate the speed.
First, we estimate the $\beta$ angle for each sample received power using minimum square error (MSE). Then, we use the linear LS estimation method to estimate the angular velocity. Then, we replace all the variables in~\eqref{eq:Lambertian_Channel_Proof2} in terms of $\beta$.

Assuming $R_o$ and $d_o$ are zero, i.e., the ViLDAR is at the end of the curvature of the road. Using some basic trigonometry identities we have  

\begin{equation}
R= r_c \sin(\beta),
\end{equation}
\begin{equation}
d_c= r_c(1-\cos(\beta)),  
\end{equation}
\begin{equation}
D=\sqrt{(d_c+d_o)^2 +(R+R_o)^2}.
\end{equation}
Substituting $h$ and $D$, we have:
\begin{equation}
D =\sqrt{r_c^2 (1-\cos(\beta))^2 + r_c^2 \sin(\beta)^2},
\end{equation}
Then,
\begin{equation}
D = r_c \sqrt{2 - 2\cos(\beta)} = 2r_c \sin(\beta/2).
\end{equation}
Given that $\cos(\theta) = \frac{R}{D}$, 
then, 
\begin{equation}
\cos(\theta) =  \frac{r_c \sin(\beta)}{2r_c \sin(\beta/2)} = \cos(\beta/2).
\end{equation}

Substituting $D(t)$ with $2r_c \sin(\beta/2)$ and $\cos(\theta(t))$ with $\cos(\beta/2)$, where $\beta$ is also changing with time, we get
 \begin{equation}
  \label{eq:Curved_senario_Lambertian_Channel}
                  P_r(\beta)=\frac{K \left ( \cos(\beta/2) \right )^{n+1}}{\left ( 2r_c \sin(\beta/2)\right )^\gamma}.
\end{equation}

To estimate the $\beta(t)$ for each measurement of $P_r(t)$, we minimize the cost function $g(\beta)$ where we define $g(\beta)$ as 
  \begin{equation}
  \label{eq:Cost_Funcation}
                  g(\beta) = \left (  P_{r,sim} -  P_{r}(\beta) \right ) ^ 2.
\end{equation}
Substituting $P_{r}(\beta)$ in ~\eqref{eq:Curved_senario_Lambertian_Channel}, we get
 \begin{equation}
  \label{eq:Cost_Funcation2}
                  g(\beta) = \left (  P_{r,sim} - \frac{K \left ( \cos(\beta/2) \right )^{n+1}}{\left ( 2r_c \sin(\beta/2)\right )^\gamma} \right ) ^ 2.
\end{equation}
The next step is to estimate the angular velocity ($w$) given that $\beta = \beta_o - wt $. We estimate $w$ and $\beta_o$ by using the linear LS equation used in~\eqref{eq:Linear_Model_LS_Final} for the straight road case. Once angular velocity is estimated, it is straightforward to find the vehicle speed given the radius of curvature.

\section{Simulation Results}\label{sec:Sim_Results}
In this section, the simulation results are presented to confirm the analytical results and investigate the impact of various system parameters on the performance of ViLDAR system.


\subsection{Straight Road Scenario}

Initially, the vehicle is very far to the point that the range and the distance are almost equal. That is to say, the angle of incidence ($\theta$) is approximately zero as shown in Fig.~\ref{fig:System_Model}.a

The following parameters are used in simulating linear LS speed estimation algorithm given by \eqref{eq:Linear_Model_LS_Final}:

\begin{itemize}


 \item Estimation duration ($\Delta t_{est}$) is 0.3\!~s unless otherwise stated (duration during which the PD is taking measurements from the approaching vehicle for the speed estimation process).

 \item The distance between the ViLDAR and the vehicle line of motion ($d$) is 0.5 m.

 \item Half power angle $\phi_{1/2}$ of the vehicle's headlamp\footnote{Although the incident angle of current headlamps ranges between $0^o$ and 40$^{\circ}$, the simulation results are provided for a wider range to observe the trend of the performance improvement.} is 40$^{\circ}$.

 \item The starting range of the simulation where ViLDAR starts taking  measurements ($R_{o}$) is equal 15 m.

 \item The ViLDAR takes a new measurement every 1 ms.

 \item The channel path loss exponent ($\gamma$) and channel gain ($K_{dB}$) are respectively $1.21$ and -49.32 dB as found for simulated channel model.


\end{itemize}

\begin{figure}[t]
 \centering
 \includegraphics[clip,trim=0 0 0 30, width=0.48\textwidth]{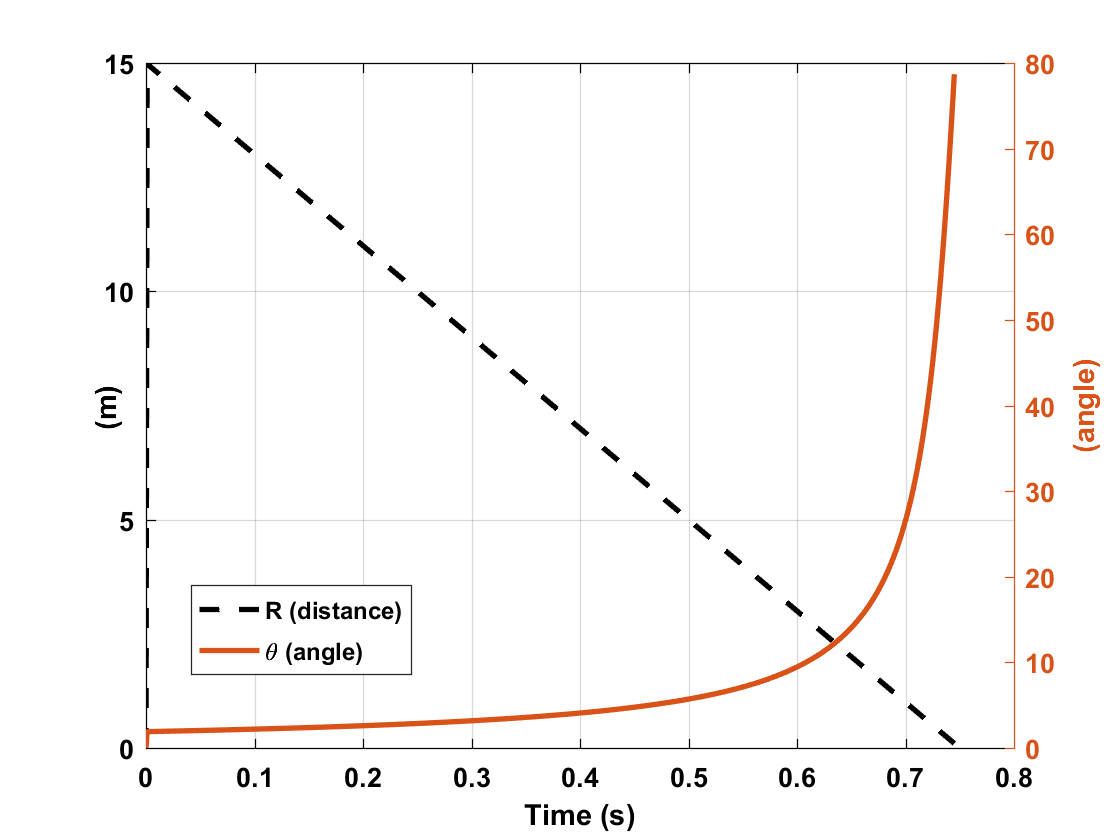}
 \caption{The incidence angle and range of the vehicle in straight road scenario assuming $R_{o}$ = 15 m, $V$ = 72 km/hr and $d$ = 0.5 m.} 
 \label{fig:Theta_Range_Vs_Time}
 \vspace{-0.3cm}
 \end{figure}

\begin{figure}[t]
\centering
\includegraphics[clip,trim=0 0 0 10, width=0.48\textwidth]{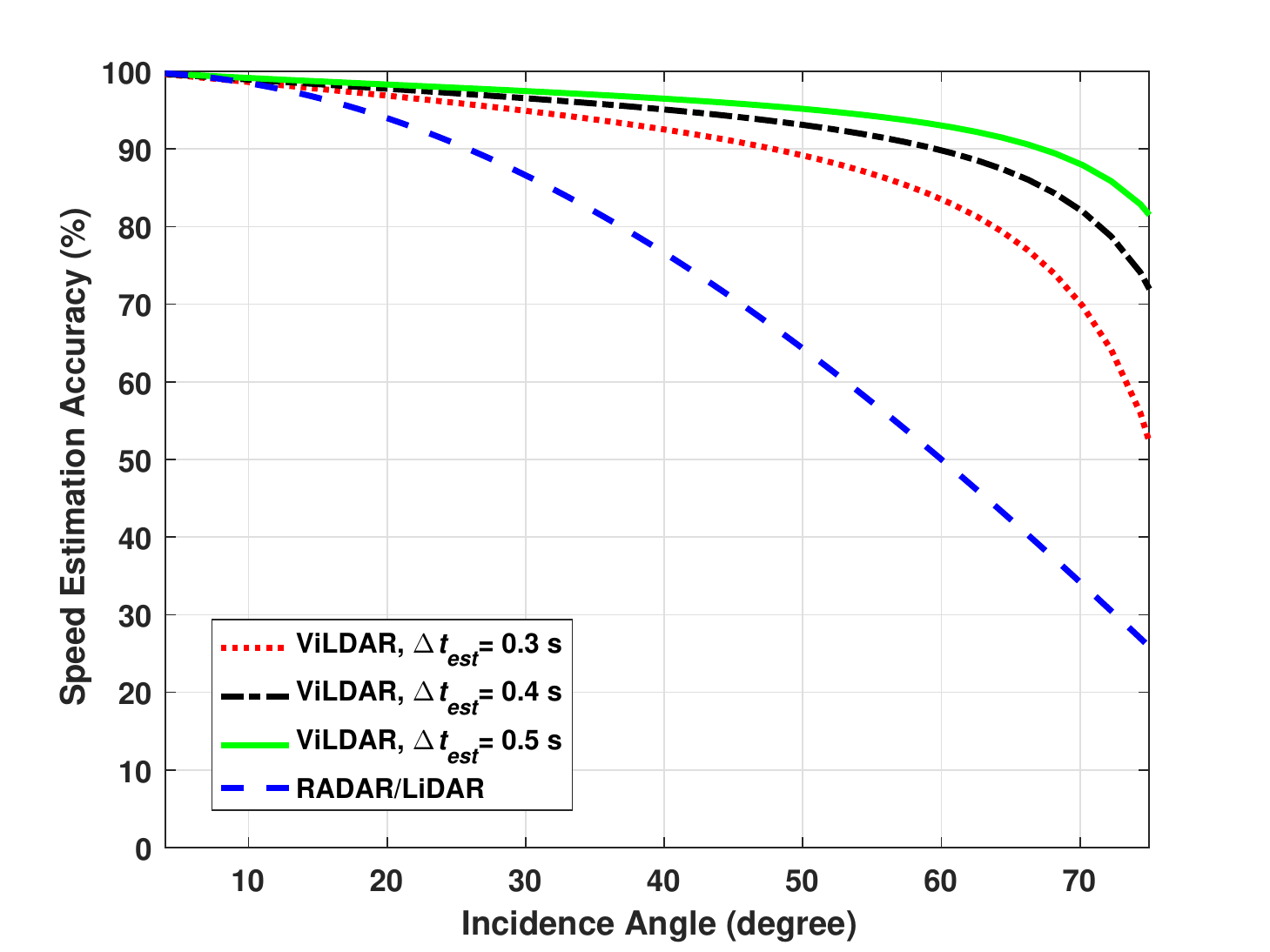}
\caption{Speed estimation accuracy of ViLDAR compared to RADAR/LiDAR for different estimation duration in straight road scenario.} 
\label{fig:RADAR_ViLDAR_Comparison}
\vspace{-0.5cm}
\end{figure}

In Fig.~\ref{fig:Theta_Range_Vs_Time}, it is shown how the the incidence angle ($\theta(t)$) and distance ($R(t)$) varies as the vehicle approaches the PD. As predicted, the range decreases and $\theta(t)$ increases as the vehicle advances toward the ViLDAR. In Fig.~\ref{fig:RADAR_ViLDAR_Comparison}, the speed estimation accuracy comparison between ViLDAR and the theoretical limit of the RADAR/LiDAR (see \cite{Doppler_RADAR_Gun_Performance2,Police_Radar_Handbook}) is presented. As it can be observed, the proposed ViLDAR system provides better performance for a wider range of incidence angles, e.g., more than \%90 estimation accuracy for up to \%80 of incidence angle range. This improvement is attributed to first ViLDAR has lower dependency on the incidence angle compared to RADAR/LiDAR systems, and second the fact that the proposed system is a one-way signal model which is less susceptible to noise and path-loss. Furthermore, as the estimation duration increases, better estimation accuracy is achieved for different incidence angles.


\begin{figure}[t]
\centering
\includegraphics[width=0.48\textwidth]{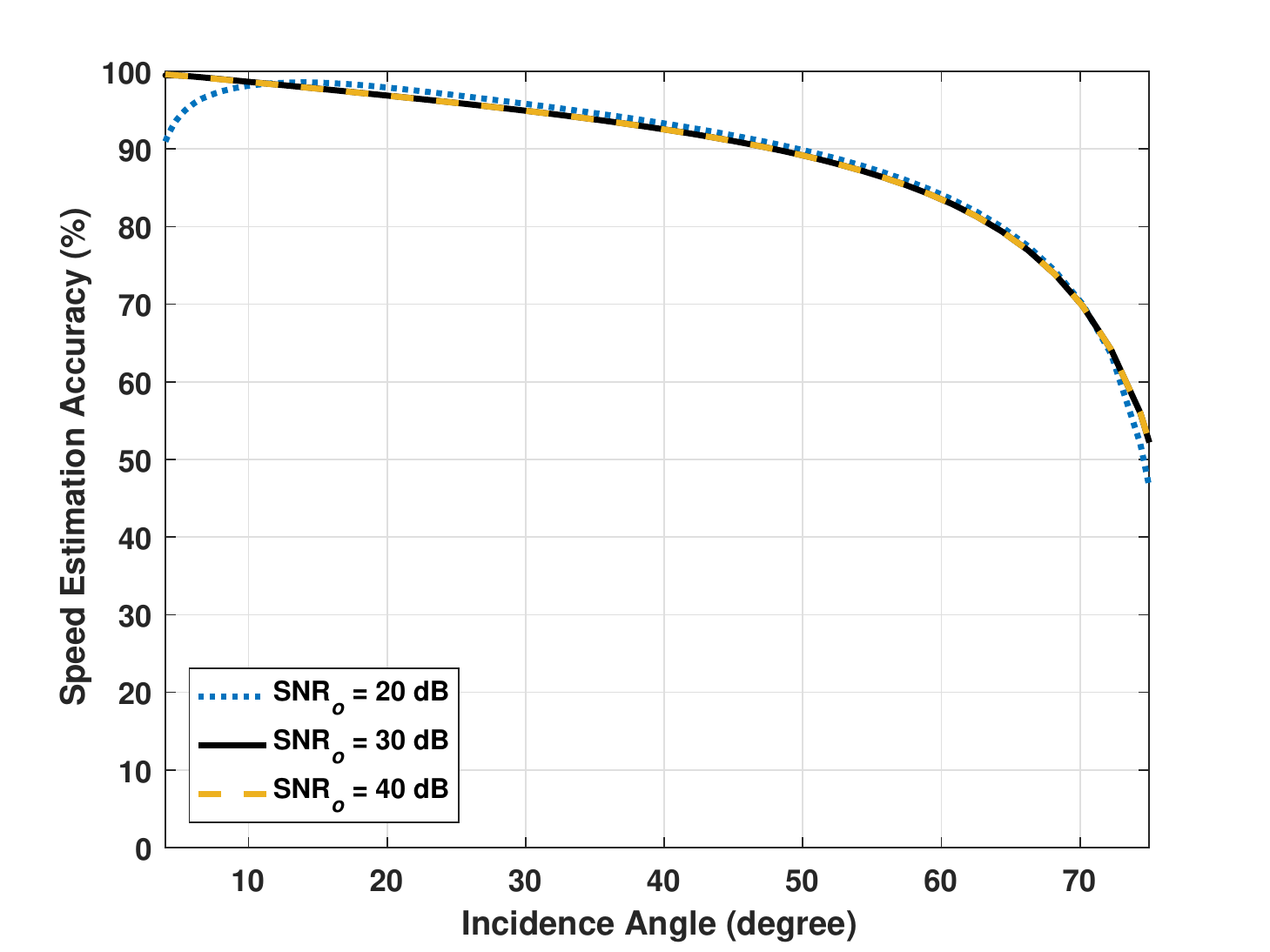}
\caption{Speed estimation accuracy for different initial SNR values in straight road scenario.} 
\label{fig:Correct_Estimation_Different_SNRs}
\vspace{-0.6cm}
\end{figure}

\begin{figure}[t]
\centering
\includegraphics[width=0.48\textwidth]{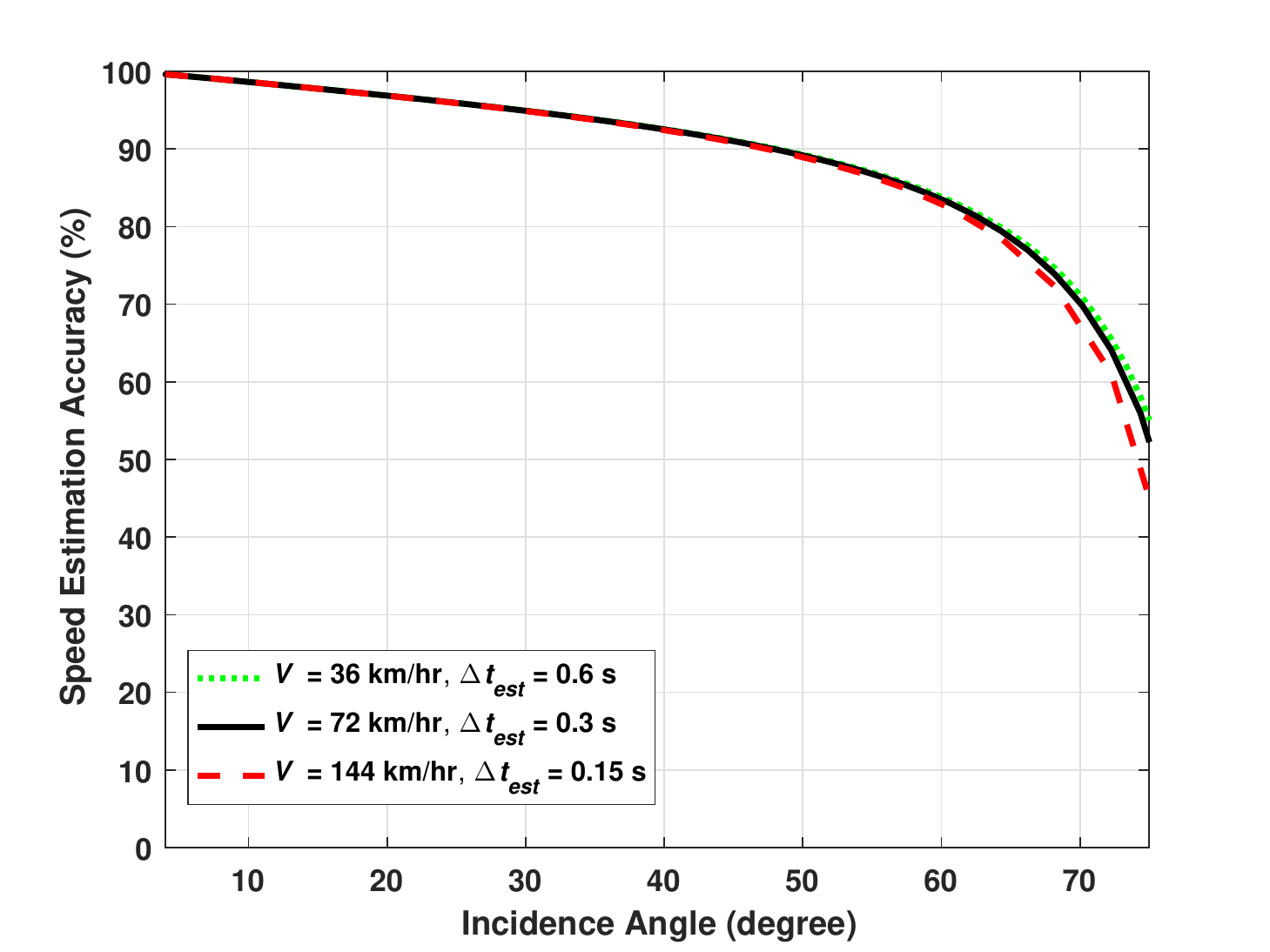}
\caption{Speed estimation accuracy of linear LS method for different actual speed values in straight road scenario.}
\label{fig:Different_Speeds_Estimation}
\vspace{-0.3cm}
\end{figure}

\begin{figure}[t]
 \centering
 \includegraphics[width=0.48\textwidth]{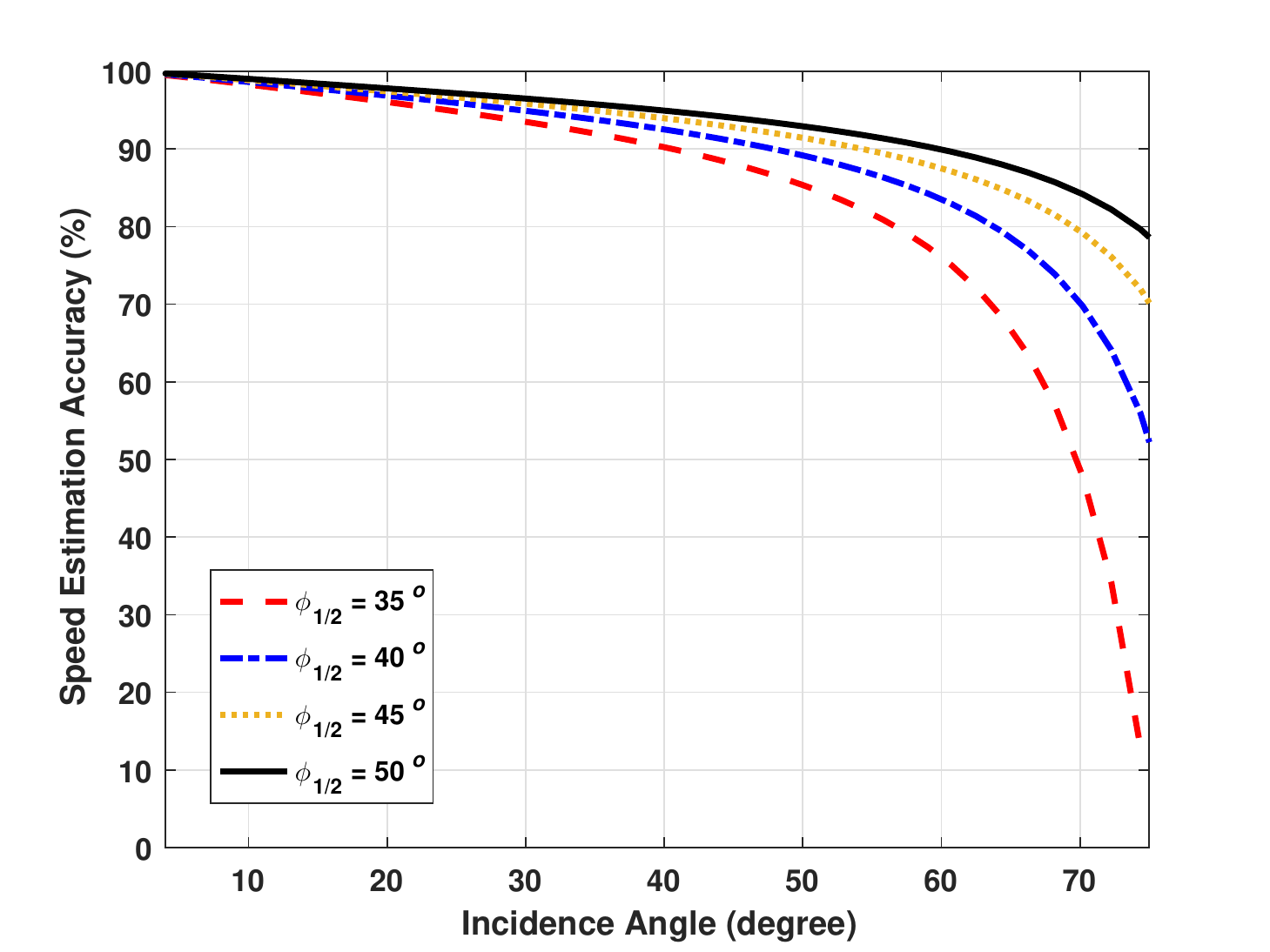}
 \caption{Speed estimation accuracy for headlamp with different half viewing angles in straight road scenario.}
 \label{fig:Different_FOVs}
 \vspace{-0.5cm}
 \end{figure}

Fig.~\ref{fig:Correct_Estimation_Different_SNRs} provides the impact of SNR levels (i.e., received power) at the initial point of measurements on the estimation accuracy. As expected, even at low incidence angles, there is performance degradation due to low SNR level. In addition, to further validate the performance of ViLDAR speed estimation method, the estimation accuracy for different speed levels and estimation duration is provided in Fig.~\ref{fig:Different_Speeds_Estimation}. Similar to the observations in Fig.~\ref{fig:RADAR_ViLDAR_Comparison}, the estimation algorithm works at different speed levels, while the performance is impacted only with the estimation duration. Moreover, as the speed of the vehicle decreases, ViLDAR needs more estimations duration (i.e., samples) to keep the same speed estimation accuracy. Finally, in Fig.~\ref{fig:Different_FOVs}, the impact of semi-half power angle ($\phi_{1/2}$) on performance is given. As expected, higher values of angles improve estimation accuracy. These results prove that the performance of the system is strongly dependent on the incident angle, number of samples used in the estimation and the noise level (accuracy) of the received power.

\begin{figure}[t]
 \centering
 \includegraphics[width=0.48\textwidth]{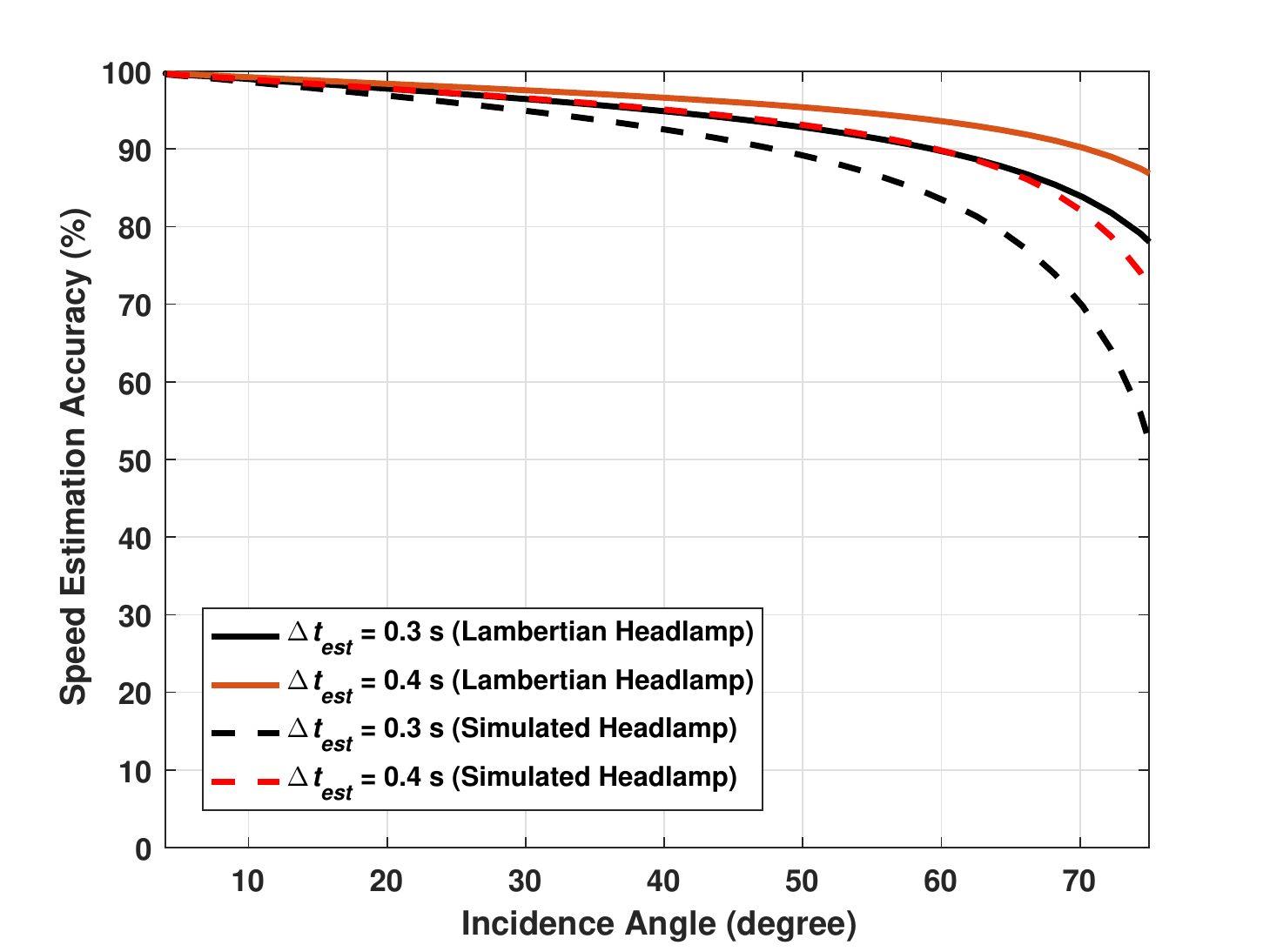}
 \caption{Speed estimation accuracy of Lambertian and simulated channel models for different estimation duration in straight road scenario.}
 \label{fig:Different_Channel_environment}
 \vspace{-0.5cm}
 \end{figure}

 In Fig.~\ref{fig:Different_Channel_environment}, we present the effect of the different channel models (simulated and Lambertian) and estimation time on the speed estimation accuracy. As expected, the performance is impacted more in simulated channel model since its channel gain ($K_{dB}$) is less than that one in Lambertian channel model.

\vspace{-0.25cm}
\subsection{Curved Road Scenario}
Initially, the vehicle is at the beginning of the curved road as shown in Fig.~\ref{fig:System_Model}.b. The following parameters are used in the simulations:

\begin{itemize}

 \item Angular vehicle speed $w$ is 1 rad/sec unless otherwise stated, where curvature radius $r_c$ is 40 m.

 \item $R_o$ and $d_o$ are zero, i.e, the ViLDAR is at the end of the curvature of the road.

 \item Half power angle $\phi_{1/2}$ of the vehicle's headlamp is 40$^{\circ}$.

 \item The ViLDAR measures the power starting at $\beta = \pi/2 $.

 \item Every 1 ms, the ViLDAR captures a new measurement.

 \item The channel path loss exponent ($\gamma$) and channel gain ($K_{dB}$) are respectively $1.21$ and -49.32 dB.

\end{itemize}

\begin{figure}[t]
 \vspace{-0.25cm}
 \centering
 \includegraphics[width=0.48\textwidth]{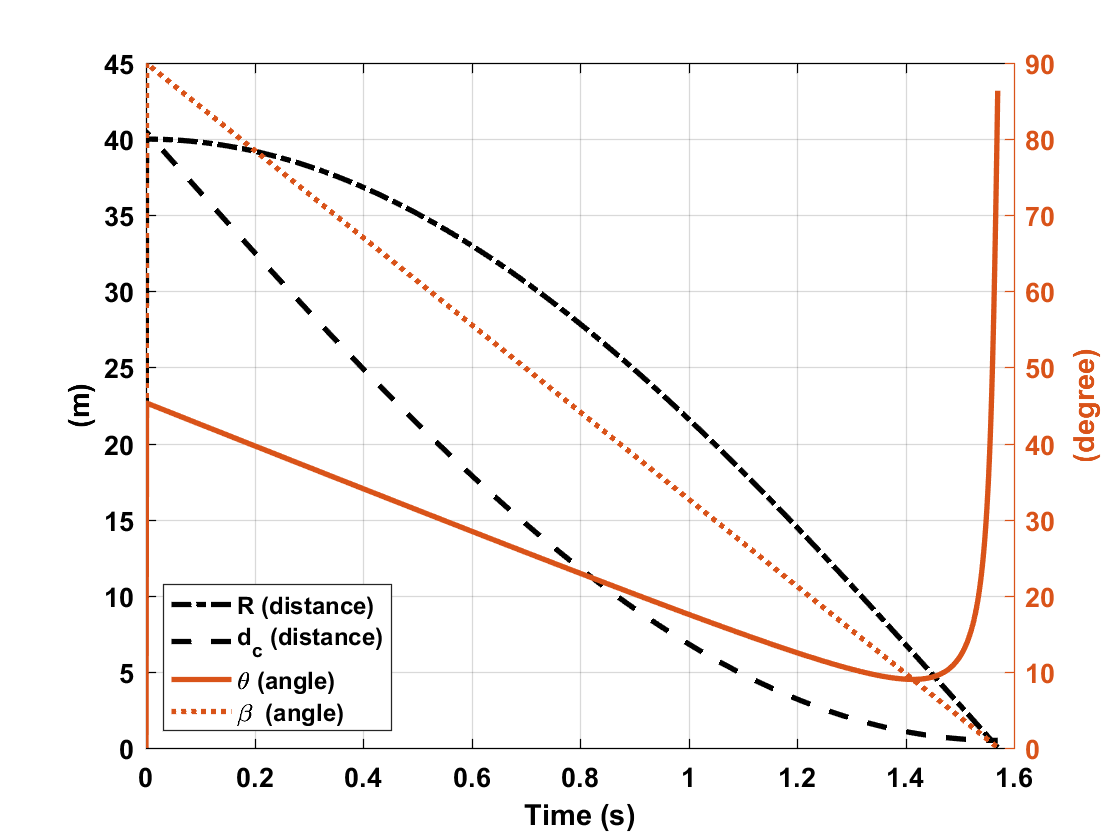}
 \caption{The incidence angle, vertical distance and range of the vehicle in curve road scenario assuming $r_c=$ 40 m and $w$ = 1 rad/sec.} 
 \label{fig:Theta_Range_Vs_Time_Curve}
  \vspace{-0.4cm}
 \end{figure}

 In Fig.~\ref{fig:Theta_Range_Vs_Time_Curve},  the change in the incidence angle ($\theta(t)$), range ($R(t)$) and vertical distance ($d_c(t)$) with simulation time is non-linear unlike the case of straight road scenario. As time increases the vehicle approaches the ViLDAR; therefore, the $R(t)$ and $d_c(t)$ decrease. While $\theta(t)$ decreases initially, then at a certain point it increases exponentially until it reaches almost $\pi/2$ due to the road curvature.

\begin{figure}[t]
 \centering
 \includegraphics[width=0.48\textwidth]{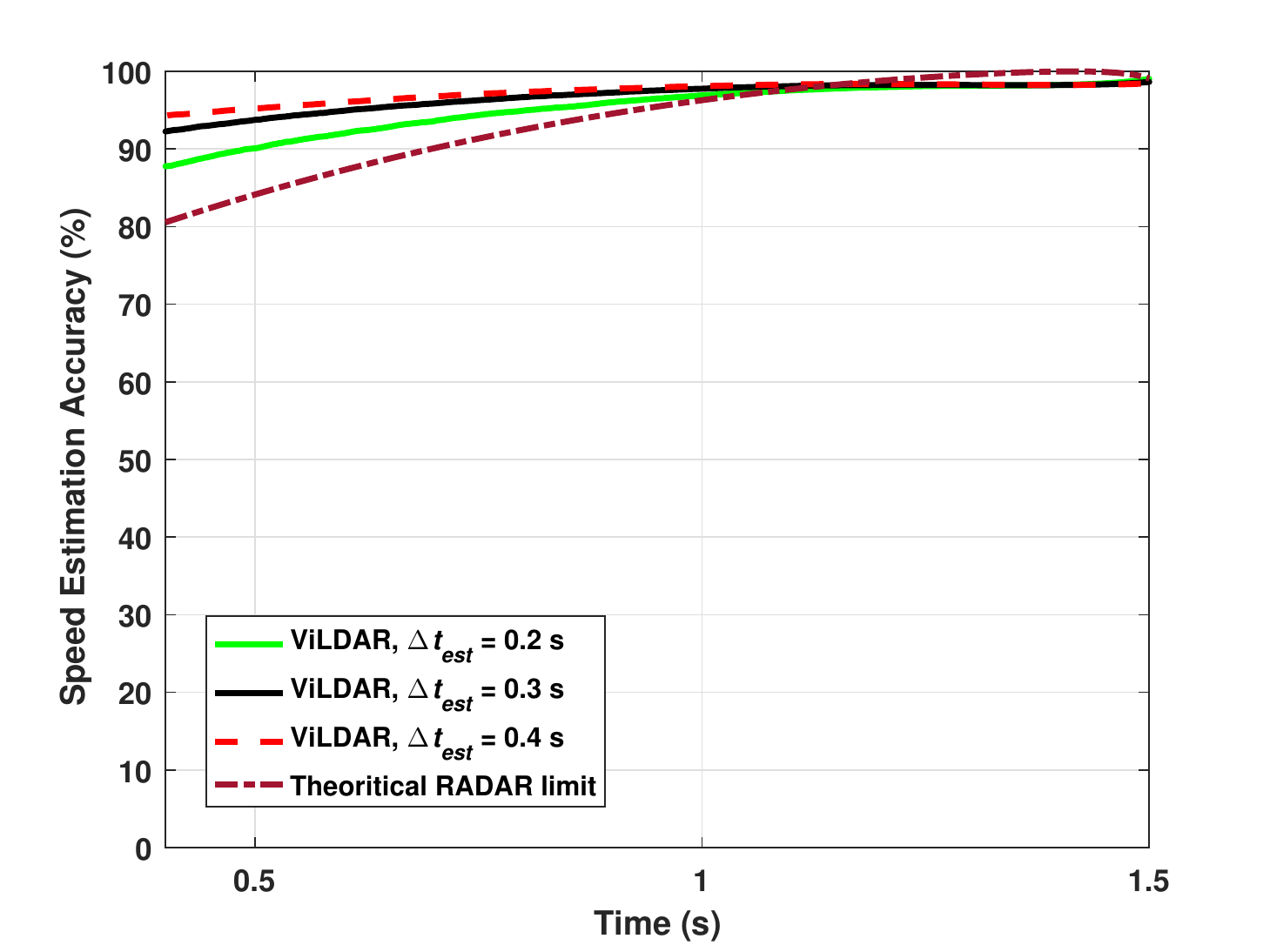}
 \caption{Speed estimation accuracy of ViLDAR compared to RADAR/LiDAR for different estimation duration in curved road scenario.}
 \label{fig:Curve_Line_Simulations_Diff_SNRs}
 \vspace{-0.4cm}
 \end{figure}

\begin{figure}[t]
 \centering
 \includegraphics[width=0.48\textwidth]{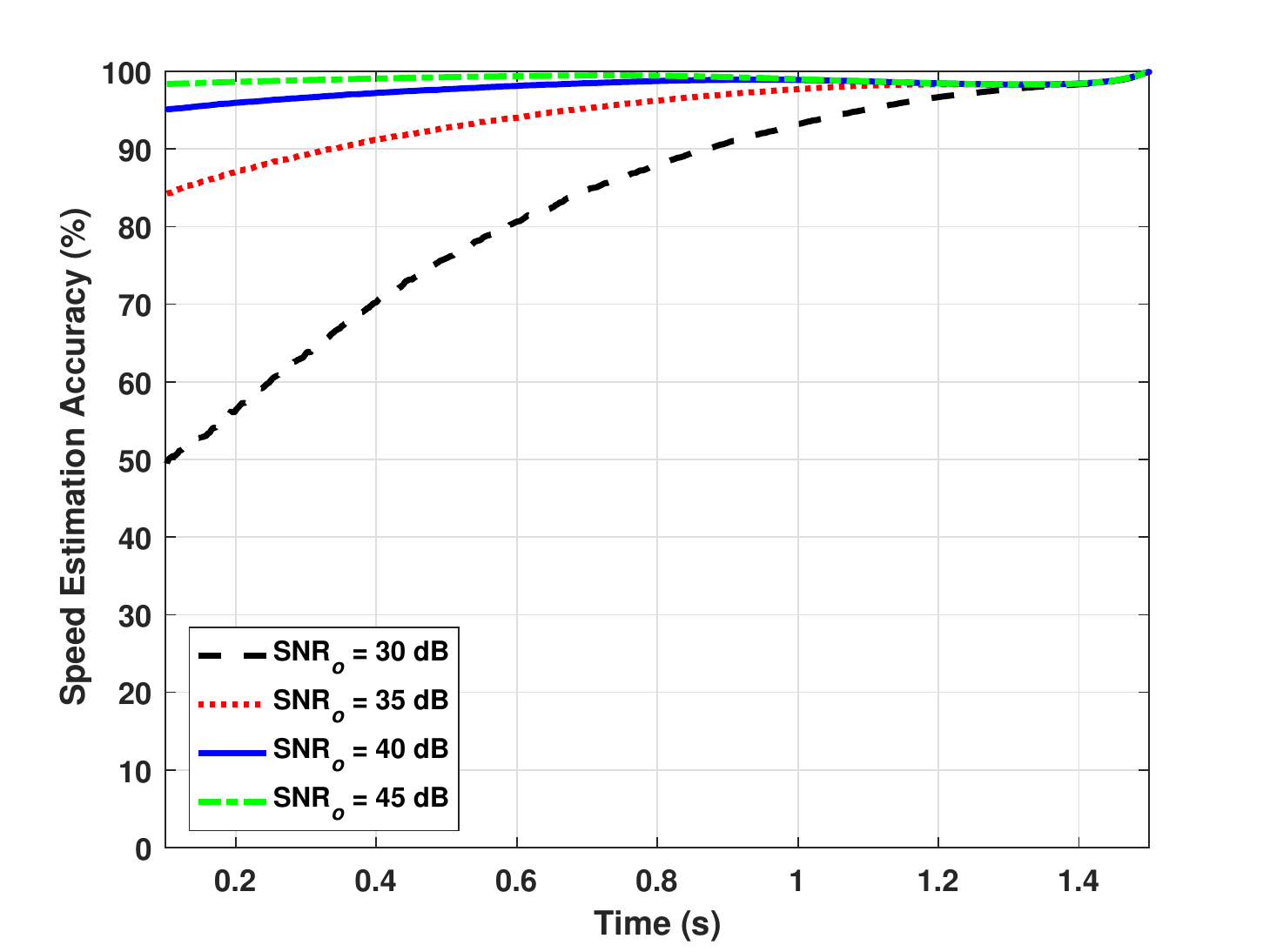}
 \caption{Speed estimation accuracy versus time for different initial SNR values in curved road scenario.}
 \label{fig:Curve_Line_Simulations_Diff_Estimation_time}
 \end{figure}

 Speed estimation accuracy is impacted by SNR levels (i.e., received power) at the initial point of measurements (see Fig.~\ref{fig:Curve_Line_Simulations_Diff_Estimation_time}) and estimation duration (see Fig.~\ref{fig:Curve_Line_Simulations_Diff_SNRs}). Notice that low performance of estimation accuracy for the initial SNR value of 30 dB in Fig.~\ref{fig:Curve_Line_Simulations_Diff_Estimation_time} is attributed to the high noise level. Moreover, as the estimation duration $(\Delta t_{est})$ increases the speed estimation accuracy improves, which can be related to benefit of having higher number of samples in estimation process.
 In addition, Fig.~\ref{fig:Curve_Line_Simulations_Diff_SNRs} presents the gain in estimation accuracy of the ViLDAR system compared to the theoretical error limit of the RADAR and LiDAR systems in curved road scenario (see \eqref{eq:Cosine_Effect_Curve} in Appendix).
\vspace{-0.2cm}
\vspace{-0.2cm}
\subsection{Comparison}
In this section, based on our simulations and discussions given in \cite{VLC_for_VehicularNetworks_DrMurat, VLC_RF_CHannel_Comparison2018}, a comparison between ViLDAR and RADAR gun is provided in Table~\ref{table:Comparison}.

\begin{table}[t]
\fontsize{11}{9}
\begin{center}
\caption{Comparison of ViLDAR and RADAR.}
\label{table:Comparison}
\resizebox{8.5cm}{!}{
\begin{tabular}{ |c|c|c| }
 \hline
 & ViLDAR & RADAR \\
\hline \hline
Range of incident angle accuracy &  High & Low \\ \hline
Range  &  Low (Up to 100 m) & High (Up to 1 km)\\ \hline
Beam-width & Wide (FOV) & Narrow \\ \hline

 Environment dependency & Sensitive & Moderate \\  \hline 
 Ambient light & Sensitive & Not affected \\  \hline
 EMI (Electro-magnetic Interference) & No & Yes \\ \hline
 
 Band license & Unlicensed & Licensed/Unlicensed \\ \hline
 Detectable by drivers & No & Yes \\ \hline
 Cost & Low & High \\ \hline
 Size & Small & Large \\ \hline 
 Power consumption & Low & High \\ \hline 
 \hline
\end{tabular}
 }
\end{center}
\vspace{-0.5cm}
\end{table}

As shown in Table~\ref{table:Comparison}, ViLDAR gives more flexibility in terms of the angle of incidence and beam-width with same high accuracy percentage. In terms of size, ViLDAR is expected to be much smaller as it only needs a PD, which can be very small in size similar to the PDs used in \cite{Study1_USING_VLC_V2V_Comm} and \cite{REAL_VLC_System_Sensor}. On the other hand, RADAR gun system needs to have the transceiver module and the antenna which depends on the frequency of operation. One of the main advantages of ViLDAR is that its presence cannot be detected by malicious drivers because of being a one-way signal model. 
Due to the fact that light wave have higher frequency than RF one used in RADARs, the operation distance range in ViLDAR will be smaller than RADAR as expected. Additional advantages and features of the ViLDAR system shared with VLC and VLS systems in general can be given as; immunity from Electromagnetic interference (EMI), using unlicensed bands and low power consumption as discussed in \cite{VLC_for_VehicularNetworks_DrMurat}.


\section{Conclusions}\label{sec:conclusion}

In this paper, a visible light sensing based speed estimation system, termed as ViLDAR, was proposed. Given the fact that the received power increases as the vehicle approaches the PD, the ViLDAR utilize linear LS method to estimate the slope of the received power with respect to time (i.e., speed). We have evaluated the performance of the proposed ViLDAR system in different road scenarios. Our results demonstrated that by using the received light intensity of vehicle's LED headlamp, the vehicle's speed can be accurately estimated for a wide range of incidence angle. In fact, more than 90\% estimation accuracy is observed for up to 70-80\% of simulation time. Comparison of results obtained for ViLDAR and RADAR also reveals that RADAR detectors poorly perform in fast incidence angle changing scenarios while promising performance can be observed for the ViLDAR system. The impact of different system parameters on speed estimation accuracy of the ViLDAR system were further investigated. It is observed that the half viewing angle of vehicle's headlamp is of crucial importance for the speed estimation accuracy.


\vspace{-0.25cm}
\section*{Acknowledgment}

The authors thank Amit Kachroo for his valuable comments and suggestions to improve this paper.
\section*{Appendix} \label{sec:RADAR_LIDAR}
In this section, the principles of RADAR/LiDAR and the factors affecting the performance of the system and the estimation accuracy are presented.

\vspace{-0.2cm}
\subsection*{Principles}

The fundamental idea of the RADAR/LiDAR is to measure the difference between the transmitted and received (after reflection) signal in frequency and time. While LiDAR systems utilize laser (light) bands, RADAR systems use RF signals. Commonly, RADAR systems are used in speed estimation in traffic control and regulations. Although, there are numerous studies that discuss how to improve detection and estimation accuracy of RADARs\cite{Speed_RF2}, there are still many limitations that impact and affect the estimation accuracy of these systems. One of the most important limitations is the LOS and narrow beam-width requirements (i.e., the angle between the device and the target) on the RADAR performance and estimation accuracy. We present a brief discussion about this challenge in the next section. Thus, the speed measurements by RADAR systems are only reliable for a certain distance, angle and availability of LOS \cite{Doppler_RADAR_Gun_Performance2}, \cite{Doppler_RADAR_Gun_Performance1}.

\subsection*{Cosine Effect}
RF- and laser-based speed detectors estimate the speed of a vehicle that is moving towards the detector.
A simplified relation between the {\it measured} speed ($V_m$) and the {\it actual} speed ($V_a$) is given as \cite{Doppler_RADAR_Gun_Performance2,Police_Radar_Handbook}:
  \begin{equation}
   \label{eq:Cosine_Effect}
     V_{m}=V_{a} \cos(\theta),
  \end{equation}
where the $\theta$, as defined in Fig.~\ref{fig:System_Model}.a, is the angle between the detector and the direction of motion of the vehicle.

Moreover, for the case of curved road scenarios\cite{Police_Radar_Handbook}, the relation can be expressed as:
   \begin{equation}
  \label{eq:Cosine_Effect_Curve}
   V_{m, curved}= V_a \sin\left\{ \frac{\pi} {2} - \beta + \tan^{-1}\frac{d_o+r_c(1-\cos\beta)}{R_o + r_c \sin \beta} \right\},
   \end{equation}
 where $\beta$ and $r_c$ are the angle and the radius of the curved-road, respectively.
 Since the angle ($\theta$) is changing fast in curved roads, microwave and laser radars cannot measure the  speed accurately. In this case, fast angle changes causes the relative speed to change too fast for the RADAR or LiDAR to measure.

If the vehicle is traveling directly towards the radar, the measured speed would be the real speed with $\theta = 0^o$. However, in practical scenarios, as shown in Fig.~\ref{fig:System_Model}, the vehicle does not travel directly towards the detector. Hence, the angle changes, which, in turn, results in an estimated speed that is different than the actual speed. This phenomenon is called the {\it Cosine Effect}, where the cosine of the angle that is between the vehicle's direction of motion and the radar is relating the calculated speed of the vehicle and the real speed. As expected, as the angle increases the estimation error increases, i.e., the detector provides less accurate results.

\bibliographystyle{IEEEtran}
\bibliography{Bibliography}

\vspace{-1.3cm}
\begin{IEEEbiography}[{ \includegraphics[width=1in,height=1.2in,clip]{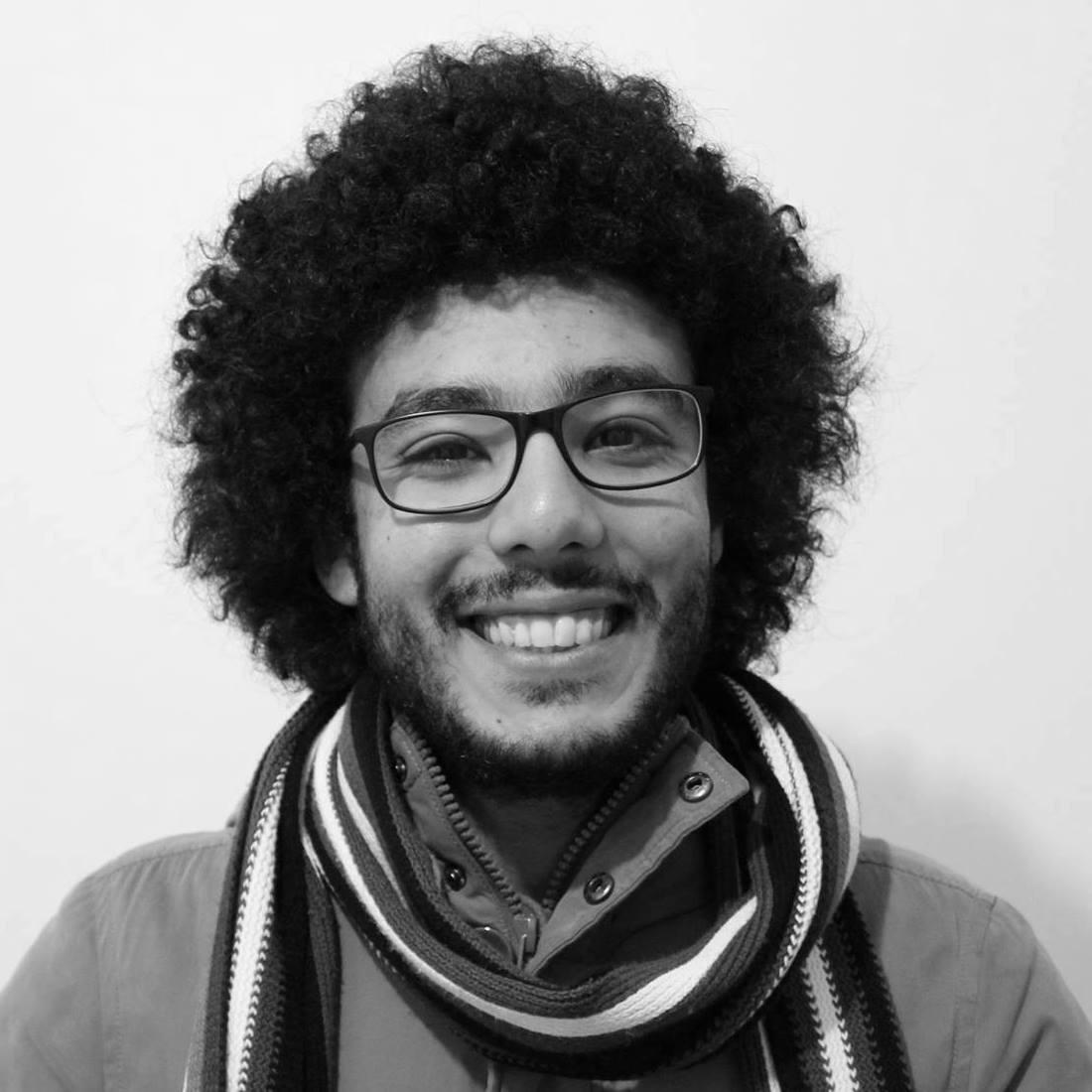}} ]{Hisham Abuella}
received the B.Sc. degree in Communications and Electronics Engineering from Ain Shams University, Cairo, Egypt, in 2013. He worked as a Digital System Design Engineer in Varkon Semiconductor Company, Cairo, Egypt for one year. In Fall 2014, he joined Istanbul Sehir University as a Research Assistant for his M.Sc. degree in Electronics and Computer Engineering at Istanbul, Turkey. Lastly, he joined Oklahoma State University as a Graduate Research Assistant to pursue his Ph.D. study at the School of Electrical and Computer Engineering in Spring 2017. He is currently working with Dr. Sabit Ekin at Wireless Communications Research Lab (WCRL). His current research interests include Visible light communication, Wireless communication systems design using SDRs , Visible light Sensing applications, and Hybrid RF/VLC systems performance analysis.
\end{IEEEbiography}
\vspace{-1.2cm}
\begin{IEEEbiography}[{ \includegraphics[width=1in,height=1.2in,clip]{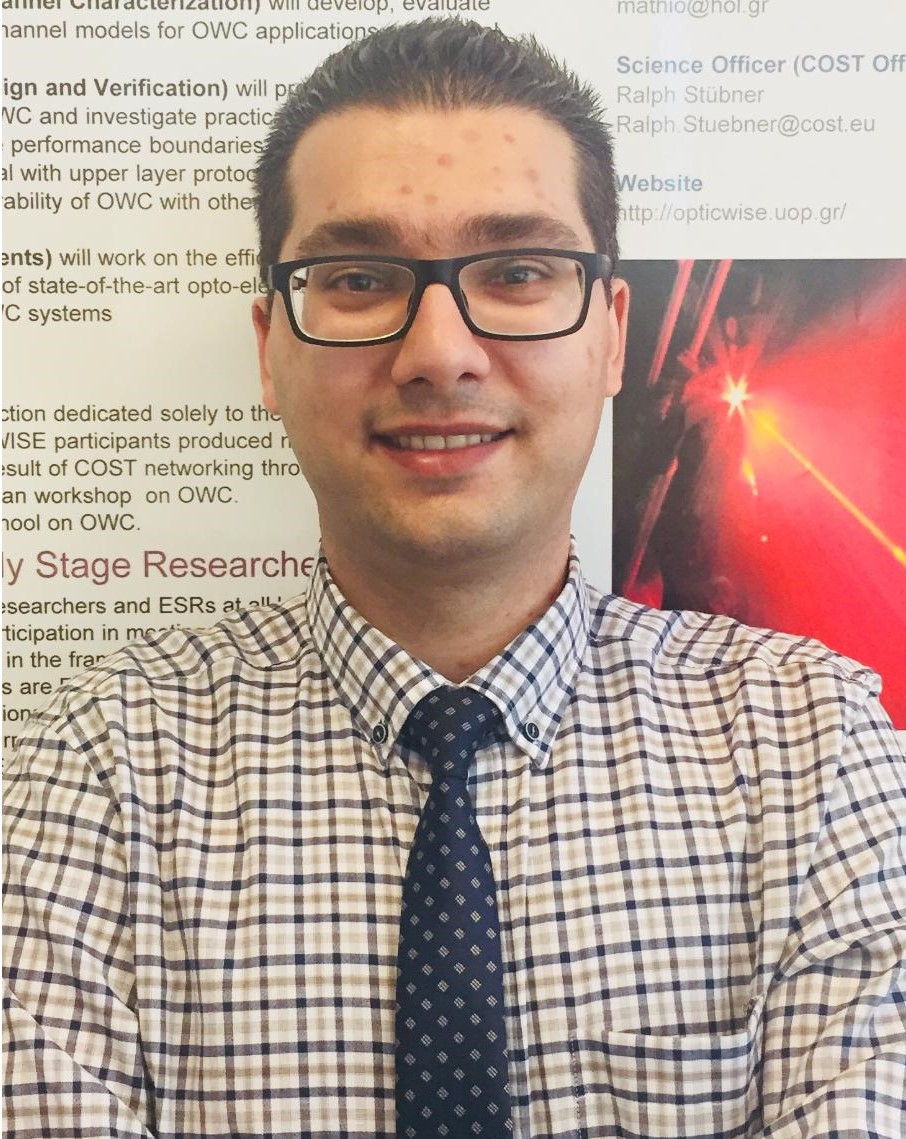}} ]{Farshad Miramirkhani}
received the B.Sc. and the M.Sc. degree with high honors in electronics and communication engineering from University of Isfahan, Isfahan, Iran, in 2011 and 2014, respectively. He joined Communication Theory and Technologies (CT\&T) Research Group as a research assistant working toward his doctorate under supervision of Prof. Murat Uysal at Ozyegin University, Istanbul, Turkey, in 2014. He will be subsequently employed by Gebze Technical University as an Assistant Professor, in 2018. He has contributed to the standardization works of IEEE 802.15.7r1 (Short Range Optical Wireless Communications). The LiFi channels developed by Prof. Murat Uysal and Mr. Miramirkhani were selected as the “LiFi Reference Channel Models” by the IEEE 802.15.7r Task Group during the IEEE’s latest meeting held in Bangkok, Thailand, in September 2015. He has served as a reviewer for several prestigious journals and conferences for IEEE and OSA societies as well as an Editorial Board Member of Optical Communications-Clausius Scientific Press. His current research interests include optical wireless communications, indoor visible light communications, underwater visible light communications, vehicular visible light communications and channel modeling. 
\end{IEEEbiography}

\begin{IEEEbiography}[{ \includegraphics[width=1in,height=1.4in,clip,keepaspectratio]{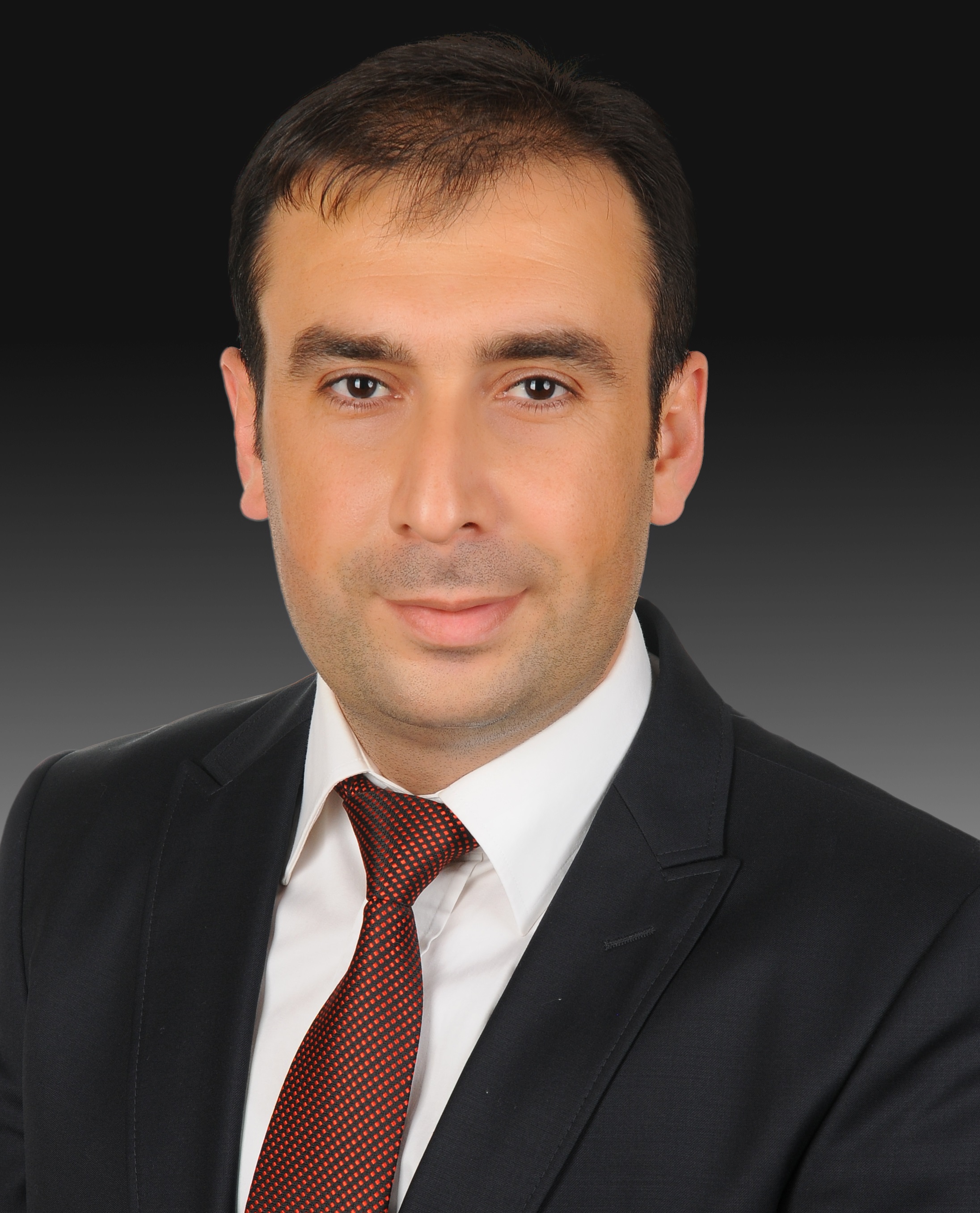}} ]{Sabit Ekin}
has joined the School of Electrical and Computer Engineering, Oklahoma State University, Stillwater, OK, as an Assistant Professor in Fall 2016. He has four years of industrial experience from Qualcomm, Inc. as a Senior Modem Systems Engineer at the Department of Qualcomm Mobile Computing. 
He received his Ph.D. from Electrical and Computer Engineering, Texas A\&M University, College Station, TX, 2012, the M.Sc. from Electrical Engineering, New Mexico Tech, NM, 2008, and the B.Sc. from Electrical and Electronics Engineering, Eskisehir Osmangazi University, Turkey, 2006.
He was working as a visiting research assistant in Electrical and Computer Engineering Program at Texas A\&M University at Qatar (2008–2009). During the summer of 2012, he worked with the Femto-cell interference management team in the Corporate R\&D at New Jersey Research Center, Qualcomm Inc. After his Ph.D. study, he joined in Qualcomm Inc., San Diego, CA, where he has received numerous Qualstar awards for his achievements/contributions on cellular modem receiver design.
His research interests are in the areas of design and performance analysis of communications systems in both theoretical and practical point of views, particularly interference management and statistical modeling of interference in next-generation wireless systems, e.g., 5G, mmWave, HetNets, visible light communications, and cognitive radio networks.
\end{IEEEbiography}
\vspace{-1cm}
\begin{IEEEbiography}[{ \includegraphics[width=1in,height=1.4in,clip,keepaspectratio]{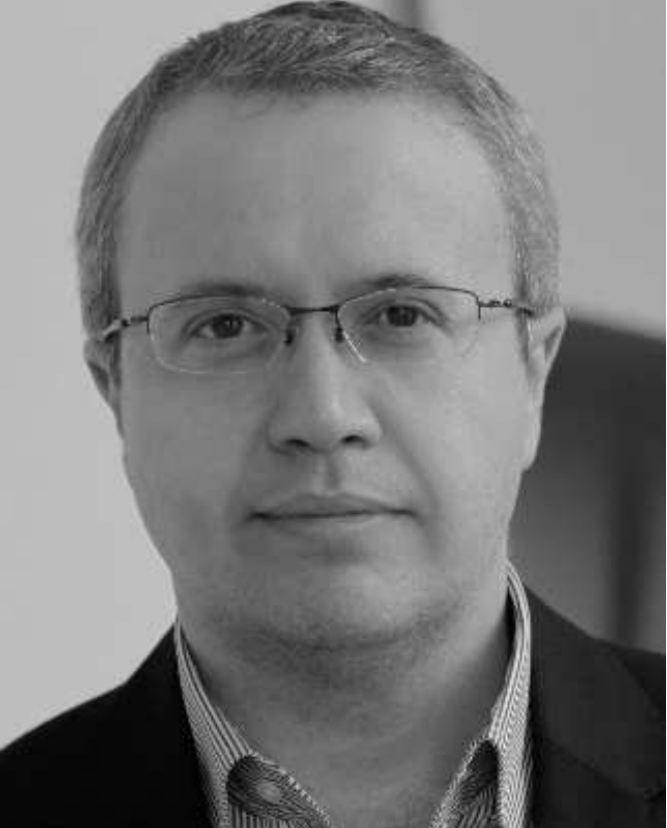}} ]{Murat Uysal}
received the B.Sc. and M.Sc. degrees in electronics and communication engineering from Istanbul Technical University, Istanbul, Turkey, in 1995 and 1998, respectively, and the Ph.D. degree in electrical engineering from Texas A\&M University, College Station, TX, USA, in 2001. He is currently a Full Professor and the Chair of the Department of Electrical and Electronics Engineering with Ozyegin University, Istanbul. He also serves as the Founding Director of the Center of Excellence in Optical Wireless Communication Technologies. Prior to joining Ozyegin University, he was a tenured Associate Professor with the University of Waterloo, Canada, where he still holds an adjunct faculty position. He has authored some 290 journal and conference papers in his research topics and received more than 7500 citations. His research interests are in the broad areas of communication theory and signal processing with a particular emphasis on the physical-layer aspects of wireless communication systems in radio and optical frequency bands. His distinctions include the Marsland Faculty Fellowship in 2004, the NSERC Discovery Accelerator Supplement Award in 2008, the University of Waterloo Engineering Research Excellence Award in 2010, the Turkish Academy of Sciences Distinguished Young Scientist Award in 2011, and the Ozyegin University Best Researcher Award in 2014. He currently serves on the editorial board of the IEEE TRANSACTIONS ON WIRELESS COMMUNICATIONS. In the past, he was an Editor of the IEEE TRANSACTIONS ON COMMUNICATIONS, the IEEE TRANSACTIONS ON VEHICULAR TECHNOLOGY, the IEEE COMMUNICATIONS LETTERS, Wireless Communications and Mobile Computing Journal, and the Transactions on Emerging Telecommunications Technologies, and a Guest Editor of the IEEE JOURNAL ON SELECTED AREAS IN COMMUNICATIONS Special Issues on Optical Wireless Communication (2009 and 2015). He was involved in the organization of several IEEE conferences at various levels. He served as the Chair of the Communication Theory Symposium of IEEE ICC 2007, the Chair of the Communications and Networking Symposium of IEEE CCECE 2008, the Chair of the Communication and Information Theory Symposium of IWCMC 2011, a TPC Co-Chair of the IEEE WCNC 2014, and the General Chair of the IEEE IWOW 2015. Over the years, he has served on the technical program committee of more than 100 international conferences and workshops in the communications area.
\end{IEEEbiography}
\vspace{4cm}

\begin{IEEEbiography}[{ \includegraphics[width=1in,height=1.4in,clip,keepaspectratio]{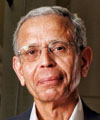}} ]{Samir Ahmed}
is a Professor of Transportation Systems \& Engineering at Oklahoma State University. He has received his B.Sc. in Civil Engineering from Cairo University, Egypt, 1968; M.Sc. in Civil Engineering from McGill University, Canada; and Ph.D. from The University of Oklahoma, USA. Dr. Ahmed has more than thirty-six years of experience in Transportation Engineering research. He has been involved in numerous research projects sponsored by the Federal Highway Administration, Federal Transit Administration, Oklahoma Department of Transportation and various transportation agencies. His research interests include: design, planning, and management of transportation systems and facilities; highway traffic operations and control; intelligent transportation/infrastructure systems; transportation safety; systems modeling, simulation, and optimization; and statistical quality assurance and quality control of highway construction.
\end{IEEEbiography}

\end{document}